\documentclass[prd,twocolumn,showpacs,preprintnumbers,showkeys,floatfix]{revtex4}
\usepackage{hyperref,amssymb,amsmath,mathrsfs,bm,graphicx,color}

\begin{document}

\title{A framework for large-scale relativistic simulations in the
characteristic approach}

\author{Roberto G\'omez}
\email{gomez@psc.edu}
\affiliation{Pittsburgh Supercomputing Center, 300 S. Craig Street, Pittsburgh,
PA 15213}
\affiliation{Department of Physics, Carnegie Mellon University,
Pittsburgh, PA 15213}

\author{Willians Barreto}
\email{wbarreto@ula.ve}
\affiliation{Centro de F\'\i sica Fundamental, Facultad de Ciencias,
Universidad de Los Andes, M\'erida, Venezuela.}

\author{Simonetta Frittelli}
\email{simo@mayu.physics.duq.edu}
\affiliation{Department of Physics, Duquesne University, Pittsburgh,
PA 15282}
\affiliation{Department of Physics and Astronomy, University of
Pittsburgh, Pittsburgh, PA 15260}

\date{\today}

\begin{abstract}

We present a new computational framework (LEO), that enables us to
carry out the very first large-scale, high-resolution computations in
the context of the characteristic approach in numerical relativity. At
the analytic level, our approach is based on a new implementation of the
``eth'' formalism, using a non-standard representation of the spin-raising
and lowering angular operators in terms of non-conformal coordinates on
the sphere; we couple this formalism to a partially first-order reduction
(in the angular variables) of the Einstein equations. The numerical
implementation of our approach supplies the basic building blocks for a
highly parallel, easily extensible numerical code. We demonstrate the
adaptability and excellent scaling of our numerical code by solving,
within our numerical framework, for a scalar field minimally coupled
to gravity (the Einstein-Klein-Gordon problem) in the 3-dimensions.
The nonlinear code is globally second-order convergent, and has been
extensively tested using as reference a calibrated code with the same
boundary--initial data and radial marching algorithm. In this context,
we show how accurately we can follow quasi-normal mode ringing. In the
linear regime, we show energy conservation for a number of initial data
sets with varying angular structure. A striking result that
arises in this context is the saturation of the flow of energy through
the Schwarzschild radius. As a final calibration check we perform a
large simulation with resolution never achieved before.

\end{abstract}



\maketitle

\section{Introduction}

The characteristic approach has been used successfully to
carry out numerical simulations of space-times with and without
sources~\cite{eth,cce,news,wobble,stable,matter,nulltube,reduced}. A
significant computational effort is nevertheless necessary to extend
its range of applicability to the simulation of astrophysically
relevant sources of gravitational radiation, such as the black
hole - neutron star binary problem, where the approach can be
most useful. As work on the characteristic formulation to date
illustrates~\cite{mode,particle,bsglrw05,matter}, it is clear that
most three-dimensional characteristic simulations, even vacuum
simulations~\cite{mode} are resolution-limited. This is particularly
true of three-dimensional simulations of systems containing compact
matter sources~\cite{particle}, even when the matter source is an
extended one. In general, all these simulations have been limited
in resolution primarily because of the time required to integrate
the equations numerically. For instance, at the finest resolution
simulation considered in~\cite{particle}, tracking a neutron star in
(close) orbit around a black hole requires approximately 1.5 months
even on one of the fastest processors currently available. This is so
even though the grid in question ($81\times 81$ angular points, $123$
points radially) is fairly moderate by today's standards. To a lesser
extent, characteristic simulations are also limited because of memory
requirements, although the characteristic scheme is particularly economic
in this regard. Even though it is feasible to equip a single-processor
workstation with the 1.4 Gbytes of memory required by that moderate grid
size, the time required for the numerical solution on even the fastest
processor would make such serial simulations highly impractical.

Most of the past code development in the characteristic
approach~\cite{news,matter} has been geared towards vector or single
processor machines. The computational platforms available today require
instead a parallel programming approach in order to perform large
resolution simulations in a reasonable time, thus a parallel version
of the characteristic code is clearly needed. In the present work we
show how, with a well thought out yet modest programming effort, it
is not only possible to produce an efficient, highly scalable parallel
implementation of characteristic codes, but to do so in such a way that
it becomes straightforward to extend our parallel implementation to new
physical models.

We aim for our numerical implementation to be particularly useful for
long--time simulations of sources of astrophysical interest, which are
very demanding in terms of the number of grid points on which to advance
the solution, and thus on the number of floating point operations
required. With that end in mind, our numerical code must scale well
in platforms with a large number of processors. We show here that our
implementation meets this goal, making it a potentially valuable tool
when applied to some of the most interesting astrophysical applications
of numerical relativity, such as the study of black hole--neutron
star binary systems, in the close orbit regime up to the tidal disruption
of the companion star.

The first astrophysical application we have in mind is a characteristic
simulation of boson stars in orbit about a black hole. For that purpose
we calibrate the current code for a massless scalar field minimally
coupled with gravitation. This let us, beyond numerical tests, compare
and calibrate our code with linear versions of it, radial codes,
and analytic (perturbative) results reported in the literature about
quasi-normal modes. The extension to a black hole--boson star system is
straightforward but not trivial, deserving a detailed study of its own and
enormous computational resources.
We want to stress that the production of gravitational waves by the
scattering of scalar waves has many mathematical features in common with
the production of gravitational waves by the motion of fluid bodies.

The article is organized as follows: In Sec.~\ref{sec:standard}
we review briefly the standard numerical implementation of the {\it
eth} approach~\cite{eth}, and discuss some of its drawbacks when
applied to high resolution simulations in the characteristic approach
to numerical relativity. In Sec.~\ref{sec:gnomic} we present an
implementation of the $\eth$ and $\bar\eth$ operators which combines
their standard description in terms of stereographic coordinates with
their numerical representation on non--conformal grid coordinates on the
sphere. As this approach differs significantly from our previous work,
we provide further motivation for this departure. Sec.~\ref{sec:numerics}
provides a detailed description of the numerical implementation of the
approach outlined in Sec.~\ref{sec:gnomic}. In Sec.~\ref{sec:ekg3d}
we illustrate how the parallel, scalable characteristic code
framework (LEO) that we have developed, can be used to implement the
model problem of a scalar field minimally coupled to gravity in three
dimensions. Sec.~\ref{sec:addnumer} expands on additional numerical
considerations specific to the hypersurface and evolution equations,
and to the boundary conditions, and presents convergence tests of our
numerical implementation. In Sec.~\ref{sec:results}, the viability of
the approach is demonstrated by clearly resolving several problems which
could not be tackled previously. We close in Sec.~\ref{sec:conclusions}
with concluding remarks and an outline of future work.

\section{The standard {\it eth} approach in characteristic numerical
relativity}

\label{sec:standard}

The characteristic approach to numerical relativity is based on
null coordinates $x^\alpha=(u,r,x^A)$, with $u$ the retarded time,
$r$ a luminosity distance and $x^A$ coordinates on the sphere~\cite
{eth,cce,news,wobble,stable,matter,nulltube,reduced,fission,quasi}.
In its 3-dimensional implementation, the angular coordinates chosen
are stereographic coordinates $x^A=(\zeta,\bar\zeta)$ on the sphere,
which is covered with two stereographic coordinate patches, as
first presented in Ref.~\cite{eth}. We summarize here the salient
aspects of the approach to provide the motivation for (and highlight
the differences with), the implementation described in this article.
The standard {\it eth} approach~\cite{eth} is a straightforward numerical
implementation in stereographic coordinates $x^A=(\zeta,\bar\zeta)$
of the $\eth$, $\bar\eth$ operators introduced by Newman and
Penrose~\cite{np,goldberg}. Two stereographic patches are used to cover
the unit sphere, with coordinates $\zeta_N=\tan(\theta/2)e^{i\phi}$ in
the north patch and $\zeta_S=1/\zeta_N$ in the south patch, respectively,
where $(\theta,\phi)$ are standard angular coordinates.

In terms of the dyad $q^A=P(1,i)$, where $P=1+\zeta\bar\zeta$, vectors
$U^A$ on the sphere are represented by a spin-weight 1 field $U=q^A
U_A$ (or alternatively, by a spin-weight -1 field $\bar U = \bar q^A
U_A$). This treatment generalizes to tensors on the sphere $T_{A\ldots
N+M}$, which are represented in terms of spin-weighted functions 
obtained by contracting them with the dyad $q^A$ and its complex conjugate $\bar
q^A$, i.e.
\begin{eqnarray}
   \Psi = q^{A_1} \ldots q^{A_N} \ \bar q^{A_{N+1}} \ldots \bar q^{A_{N+M}}\nonumber\\
   \times T_{A_{1} \ldots A_{N}\,A_{N+1} \ldots A_{N+M}}.
\end{eqnarray}
The spin of the resulting scalar function is $s=N-M$. Angular derivatives
of tensor fields are represented by the action of the spin-raising
and lowering operators $\eth$ and $\bar\eth$. For example, the angular
derivatives of a vector field $\nabla_A U_B$, where $\nabla_A$ are the
derivatives compatible with the flat sphere metric in the coordinates
$x^A$, are represented by the spin-2 field $\eth U$ and spin-0 field
$\bar\eth U$ given by
\begin{equation}
    \eth U = q^A q^B \nabla_A U_B, \quad
    \bar\eth U = \bar q^A q^B \nabla_A U_A .
\end{equation}
The $\eth$ and $\bar\eth$ operators acting on a spin-weight $s$ function
$\Psi$ are equivalently defined by
\begin{subequations}
\begin{eqnarray}
   \eth \Psi & = P^{1-s} \partial_{\bar\zeta} \left(P^s \Psi \right)
             & = (1 + \zeta \bar \zeta) \partial_{\bar\zeta} \Psi
               + s \zeta \Psi,
  \\
  \bar \eth \Psi &= P^{1+s} \partial_{\zeta} \left( P^{-s} \Psi \right)
             & = (1 + \zeta \bar \zeta) \partial_{\zeta}  \Psi
               - s \bar \zeta \Psi,
\end{eqnarray}
\label{eq:eth}
\end{subequations}
where, in terms of the (real) coordinates $(q,p)$,
$\zeta=q+ip$, $\partial_\zeta=\partial_q-i\partial_p$,
$\partial_{\bar\zeta}=\partial_q+i\partial_p$. Functions on the sphere
with spin-weight $s$ transform between patches according to
\begin{equation}
\Psi_{N} = \left(-\frac{\bar\zeta_S}{\zeta_S}\right)^s \Psi_{S}.
\end{equation}

We implement this numerically by laying down a two-dimensional grid on
each patch, with coordinates $(q_m,p_n)$, $\zeta_{m,n}=q_m+i p_n$, such
that $q_m= -1+(m-3) \Delta$, $p_n=-1+(n-3) \Delta$, $\Delta=2/(N-5)$,
and with the grid point indices in the range $m,n = 1 \ldots N$. This
grid covers the coordinate range $-1-2\Delta \le (q,p) \le 1+2\Delta$.
Ghost zones are used on each side of the grid for the discretization
of angular derivatives by centered, second-order-accurate finite
difference stencils. Function values at these ghost zones are obtained
by interpolation from the function values on the opposite patch. In
order to compute second angular derivatives to second order accuracy,
the interpolations must be evaluated to fourth-order accuracy,
which can be readily attained using a sixteen-point stencil in two
dimensions~\cite{eth}, provided the grid covers no less than the range
indicated above. For some applications~\cite{mode}, we find it necessary
to extend the grid to a finite overlap, i.e. $|q| \le 1+\epsilon$,
with $\epsilon\ge 2\Delta$.

The set of ghost zones required for the north patch, maps onto the south
patch (and vice-versa) as per the transformation $\zeta_N=1/\zeta_S$,
into a cloverleaf shape. Regardless of the number of grid points
(or of ghost zones), there is a finite overlap between the north and
south patches. While the points in the overlap area of each grid are
redundant, we carry them all because we find it more efficient to work
with rectangular grids. A potential problem of doing so is the development
of two different numerical solutions in the overlap area of each patch,
only loosely coupled at the stereographic patch edge.

We now discuss briefly some of the drawbacks of the standard ``eth''
approach when applied to high resolution simulations in the characteristic
approach to numerical relativity, and the motivation for the changes
that we propose in the next section.

\subsection{Parallelization of existing characteristic codes}

The first objection that we encounter is in the process of parallelizing
our characteristic codes. Because of the radial march implicit in the
radial integration of the hypersurface equations, the natural way to
parallelize a characteristic simulation is to distribute the angular grid
among processors, which would integrate the equations along a ``pencil''
of null rays. In the computational ``eth'' approach this means assigning the
computation of the solution over a subset of each stereographic patch to
a given processor. A similar arrangement, in the context of axisymmetric
simulations, was explored earlier by Bishop et. al.~\cite{transputer}.

Thus, given $M\times M$ processors, we can simply partition the $N\times
N$ stereographic grid on each patch, assigning equal square subgrids
of extent $N/M$ on each direction to each processor. Load-balancing
(the requirement that all processors in a parallel computation do
approximately the same amount of work) would in principle be achieved,
so long as we restricted ourselves to explicit methods, thus guaranteeing
that the time spent per subgrid remains constant. The communication
pattern imposed by the two stereographic patches implementation of the
``eth'' approach~\cite{eth} does present an obstacle to effective scaling.
The mapping of ghost zones at the edge of the grid to grid points
in the opposing patch is not restricted to nearest neighbors. To provide
values for these ghost zones requires data from a set of grid points
whose values are scattered among processors in an irregular pattern (in
the sense that, depending on its location on the angular grid, the ghost
zones may be obtained from one, two, or more processors). This procedure
is not only cumbersome to program, but intrinsically inefficient.

If the data required for these ghost zones could be obtained just from
grid points at the edge of an adjoining grid, the procedure would simplify
considerably. The time spent in communication would be substantially
reduced and remain constant over the set of processors, with a significant
impact on the scalability and overall efficiency of a code. Unfortunately
such an arrangement is not possible with a stereographic grid. In
addition, as we point out at the end of Sec.~\ref{sec:standard},
because of the fixed overlap between patches, a significant portion
of the grid is wasted. While this might be acceptable in small scale
simulations~\cite{matter,particle,mode,bsglrw05}, it needs to
be addressed in the context of large scale computations as in that case
it translates into a serious waste of computational resources.

Yet another problem that arises from the angular grid layout is that of
highly non-uniform angular resolution, as a direct consequence of using a
stereographic grid. Considering the expression for the area element in
stereographic coordinates,
\begin{equation}
   ds^2 = \frac{4}{(1+\zeta\bar\zeta)^2} \,d\zeta d\bar\zeta ,
\end{equation}
it can be seen that there is a marked disparity between the resolution
at grid points at various places on the sphere. Considerable better
resolution is attained near the equator than at the poles, with the
largest disparity between a grid zone at the pole ($\zeta\bar\zeta=0
$) and a point at a corner of the grid ($\zeta \bar\zeta=2$) where the
respective area elements have a ratio of 9:1. For some situations, such
as the case of a matter source in equatorial orbit around a black hole,
this feature works to our advantage. Conversely, for a matter source in
a polar orbit, the matter source would be resolved three times as poorly
when it lies along the $z$ axis ($q=p=0$) compared to the resolution
obtained when it crosses the equator, which in a second-order accurate
code translates into a nine-fold increase in the intrinsic error in the
numerical solution. One possible correction for this effect would be
to maintain the standard stereographic grid as the basic computational
grid, but introduce a physical grid related to the computational grid
by a fish-eye stretch in the angular coordinates, and allow the refined
portion of the grid to follow the compact object.

A related issue is that of proper resolution of angular features, which is
critical for characteristic simulations of compact objects in orbit around
a black hole. In spherical coordinates, angular resolution decreases
with distance to the center, as pointed out in~\cite{particle}. There is
then a limit to the distance at which we can initialize a characteristic
simulation, a limit which is not dictated by physical considerations,
such as the need to avoid the formation of caustics, which would lead to
a break down of the coordinate system. We are also limited by the number
of points which can in practice be devoted to resolve a companion object.

A proposed extension to the characteristic approach~\cite{particle}
would include an adaptive mesh refinement (AMR) strategy, which
indirectly would address both grid resolution issues mentioned. While
this extension may prove necessary in simulations of the last stages
of capture or disruption of a companion star by a black hole, the
necessary technology has not yet been developed in the context of
the characteristic approach; to our knowledge, applications of AMR in
the characteristic framework have been made only in simplified one-
dimensional models~\cite{nullamr}. The intermediate stages of the black
hole--neutron star problem, where the companion remains approximately
contained in a finite region could, at least in principle, be equally
well resolved with fixed mesh refinement, an approach successfully used,
for example, in~\cite{schnetter,sperhake,imbiriba}. The introduction of
AMR carries with it a whole new set of issues, not the least of which is
the problem of load-balancing in a massively parallel computer. Current
characteristic codes present serious obstacles to the implementation of
an AMR strategy by the nature of the angular grid alone. To be effective,
an AMR implementation would have to be capable of dealing with refined
meshes in multiple coordinate patches. To our knowledge, few existing
implementations have this capability~\cite{Overture}, and none have
been applied in numerical relativity. In the present work we present a
parallel implementation on a single distributed grid, and we defer the
discussion of possible techniques for fixed and adaptive mesh refinement
for future work.

\section{An $\eth$ operator based on non-conformal projections}
\label{sec:gnomic}

A key consideration for the present work is that different numerical
representations of the $eth$ approach can be developed by laying down
different types of grids on the sphere. We consider here an alternative to
the standard ``eth'' approach, based on the ``cubed sphere'' or ``gnomic''
covering of the sphere introduced by Ronchi {\it et al.}~\cite{ronchi}, based on
earlier work of Sadourny~\cite{sadourny}. This approach is now in common
use in global weather simulations, such as in the MIT General Circulation
Model (MITgcm), for example~\cite{adcroft}. It has also been applied in
astrophysical simulations~\cite{koldoba,romanova}, in the study of wave
propagation methods on the sphere~\cite{rossmanith}, and more recently,
in the evolution of scalar fields in a fixed background~\cite{lrt}. While
preparing this manuscript, we learned~\cite{nigel} of work being carried
out by Bishop {\it et al.}~\cite{bishop}, on a similar grid arrangement,
based on work by Thornburg~\cite{thornburg}.
\begin{figure}[!ht]
\includegraphics[width=3.25in,height=3.25in,angle=0]{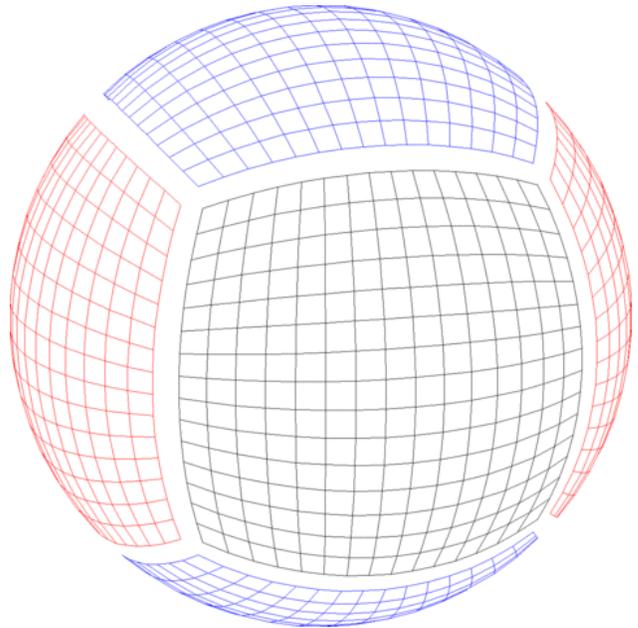}
\caption{The cubed--sphere: covering of the sphere with six
non--overlapping gnomic patches.}
\label{fig:gnomic}
\end{figure}

In the `cubed sphere' method, a covering of the unit sphere with six
non-overlapping patches results from projecting the sphere from its center
onto the six faces of a circumscribing cube, whose edges
have length two.  For example, for points on the sphere with Cartesian
coordinates $ (x,y,z)$ and angular coordinate $-\pi/4 \le \theta\le
\pi/4$ (where $z=\cos\theta$), we project the point by tracing a line
from the center of the sphere though the point $(x,y,z)$ to the $z=1$
face of the circumscribing cube, determining a point with Cartesian
coordinates $(U,V,1)$, we then label the point on the sphere according
to the Cartesian coordinates $(U,V)$ of its projection on the plane $z=1$.

In order to obtain a covering of the sphere with nearly uniform
area, we label the points on the sphere by angular coordinates
$(\alpha,\beta)$, where $ U=\tan(\alpha)$, $V=\tan(\beta)$, introducing
an equally spaced grid in the angular coordinates $(\alpha,\beta)$,
i.e. $(\alpha_i,\beta_j)=(i \Delta_\zeta, j \Delta_\zeta)$, $i,j=-
N...N$, $\Delta_\zeta=\pi/(4N)$. Similar projections from the center
of the sphere to the other faces of the cube provide a covering of the
sphere with six patches.

For a finite-difference code, this arrangement is ideal, as the grid
on each patch is equally spaced in the angular coordinates, and the
same angular coordinate is used on any two patches in the direction
perpendicular to a boundary. Thus any additional layers of ghost zones
in the adjacent grid will fall on coordinate lines parallel to the
boundary. Evaluating function values at those ghost zones requires only
one--dimensional interpolation along coordinate lines parallel to the
boundary. For example, an $N$-th order centered stencil requires $N/2$
additional layers of ghost cells to be supplied along the edge of each
spherical cap. The order of the interpolations used to supply these
points can be selected so as to preserve the accuracy and maintain the
desired dissipation properties of the numerical scheme. In the standard
configuration~\cite{ronchi}, the spherical caps share common points along
the edges of the grid, where two grids abut, and at the corners of each
patch, where three grids meet. These common points must have a unique
value on each of the grids that share them. An approach advocated in
the literature~\cite{rossmanith} is to replace the function values at
these shared points by some form of weighted average. Here we dispense
with this procedure, by carefully selecting the range of the angular
coordinates in each patch, in a way that precludes the existence of
points common to two (or three) cubed--sphere patches.

An advantage of a {\it gnomic} decomposition of the sphere, crucial to
an efficient parallel implementation of the numerical $\eth$ approach,
is that the patches can be laid out so that they are non-overlapping. In
contrast, in a stereographic covering, there is a finite overlap, which
does not shrink in size as we increase the grid resolution. Although
it is possible to construct a six--patch stereographic covering of the
sphere, in which the amount of overlap is reduced with respect to the
two--patch covering, the area of the overlap zone remains constant,
regardless of grid size.

\subsection{Non-conformal projections of the sphere}

Non-conformal projections are based on non-orthogonal coordinates,
thus there is no symmetry in some operators such as the Laplacian on
the sphere. It is still straightforward to couple a gnomic grid layout
with the existing $\eth$ approach, i.e. while continuing to express the
$\eth$ and $\bar\eth$ operators of Ref.~\cite{eth} on stereographic
coordinates. In the following we detail how this is implemented by
expressing the angular (stereographic) derivatives in terms of angular
derivatives in gnomic coordinates. A gnomic covering of the sphere is
given by six coordinate patches
\begin{subequations}
\begin{eqnarray}
\left( x_1, y_1, z_1 \right) &=& \frac{1}{D_1} \left( 1,    U_1,  V_1  
\right),
\\
\left( x_2, y_2, z_2 \right) &=& \frac{1}{D_2} \left( U_2,  1,    V_2  
\right),
\\
\left( x_3, y_3, z_3 \right) &=& \frac{1}{D_3} \left(-1,   -U_3,  V_3  
\right),
\\
\left( x_4, y_4, z_4 \right) &=& \frac{1}{D_4} \left( U_4, -1,    V_4  
\right),
\\
\left( x_5, y_5, z_5 \right) &=& \frac{1}{D_5} \left(-V_5,  U_5,  1  
\right),
\\
\left( x_6, y_6, z_6 \right) &=& \frac{1}{D_6} \left( V_6,  U_6, -1  
\right),
\end{eqnarray}
\end{subequations}
\noindent where $(x_i, y_i, z_i)$, $i=1\ldots 6$ are the Cartesian coordinates
of the points, $(U_i,V_i)$ are coordinates on the sphere in the range
$-\sqrt{2}-1 \le U_i, V_i \le \sqrt{2}-1$, and $D_i=\sqrt{1+U_i^2
+V_i^2}$. (Numerical grid points can be set equally spaced
in the coordinates $(\alpha, \beta)$, related to $(U_i,V_i)$
by $U_i=\tan(\alpha_i)$, $V_i=\tan(\beta_i)$ with $-\pi/4 \le
\alpha_i,\beta_i \le \pi/4$). The coordinate lines at $U_i=const.$
($V_i=const.$) are great circles on the sphere which pass through the
points where the Cartesian axis intersect the sphere. For instance,
the lines at $U_3=const.$ are great circles spun around the $x$ axis,
while those of $V_3=const.$ are great circles rotated around the $y$
axis, with $\alpha_3,\beta_3$ the respective rotation angles. Similarly,
on each patch, we define stereographic coordinates $\zeta_i= u_i +i\,v_i$,
where the angular coordinates $(u_i,v_i)$ on each patch are related to
Cartesian coordinates by
\begin{subequations}
\begin{eqnarray}
\left( x_1, y_1, z_1 \right) &=& \frac{1}{P_1} \left(2-P_1, 2\,u_1, 2 
\,v_1 \right),
\\
\left( x_2, y_2, z_2 \right) &=& \frac{1}{P_2} \left(-2\,u_2, 2-P_2, 2 
\,v_2 \right),
\\
\left( x_3, y_3, z_3 \right) &=& \frac{1}{P_3} \left(-2+P_3, -2\,u_3,  
2\,v_3 \right),
\\
\left( x_4, y_4, z_4 \right) &=& \frac{1}{P_4} \left(2\,u_4, -2+P_4, 2 
\,v_4 \right),
\\
\left( x_5, y_5, z_5 \right) &=& \frac{1}{P_5} \left(-2\,v_5, 2\,u_5,  
2-P_5 \right),
\\
\left( x_6, y_6, z_6 \right) &=& \frac{1}{P_6} \left( 2\,v_6, 2\,u_6  
-2+P_6 \right),
\end{eqnarray}
\end{subequations}
with $P_i=1+u_i^2+v_i^2$.  The gnomic coordinates $(U,V)$ and
stereographic coordinates $(u,v)$ on each patch are related by
\begin{equation}
    U = \frac{2 u}{1-u^2-v^2}, \quad V = \frac{2 v}{1-u^2-v^2} ,
\end{equation}
or, in more compact form, in terms of a complex gnomic coordinate $\xi
= U+i\,V$,
\begin{equation}
   \xi = \frac{2 \zeta}{1 - \zeta \bar \zeta} , \quad
   \zeta = \frac{\xi}{1+\sqrt{1+\xi\bar\xi}}.
\label{eq:xi_zeta}
\end{equation}

\subsection{The eth operator on the cubed sphere}

We can express the angular derivatives in stereographic coordinates,
$\partial_{\zeta}$, $\partial_{\bar\zeta}$ that enter into the $\eth$
and $\bar\eth$ operators in terms of angular derivatives $(u,v)$ through
the relations
\begin{equation}
  \frac{\partial}{\partial\zeta}  =\frac{1}{2} \left(
          \frac{\partial}{\partial u}
      - i \frac{\partial}{\partial v} \right)\ , \quad
  \frac{\partial}{\partial\bar\zeta} = \frac{1}{2} \left(
          \frac{\partial}{\partial u}
      + i \frac{\partial}{\partial v} \right)
\end{equation}
and with the Jacobian
\begin{eqnarray}
\left(
  \begin{array}{ccc}
     \displaystyle{\frac{\partial U}{\partial u}} &
     \displaystyle{\frac{\partial U}{\partial v}} \\
     & \\
     \displaystyle{\frac{\partial V}{\partial u}} &
     \displaystyle{\frac{\partial V}{\partial v}} \\
  \end{array}
\right) &=&
   \frac{2}{(1-u^2-v^2)^2} \times \nonumber \\
   &&
    \left(
        \begin{array}{ccc}
              1+u^2-v^2 &  2 u v     \\
                        &            \\
              2 u v     &  1-u^2+v^2 \\
        \end{array}
    \right)\,
\end{eqnarray}
we cast derivatives with respect to stereographic $(u,v)$ in terms of
derivatives with respect to gnomic coordinates $(U,V)$, which are in
turn related to derivatives with respect to gnomic coordinates
$(\alpha,\beta)$, by
\begin{equation}
    \frac{\partial}{\partial U} = \frac{1}{(1+U^2)}
                                    \frac{\partial}{\partial  
\alpha} , \quad
    \frac{\partial}{\partial V} = \frac{1}{(1+V^2)}
                                    \frac{\partial}{\partial \beta} ,
\label{eq:dbydU}
\end{equation}
The spin-raising and lowering operators $\eth$ and
$\bar\eth$~\cite{stewart} acting on a spin $s$ function $\Psi$,
Eq.~(\ref{eq:eth}), can be written as
\begin{eqnarray}
  \eth \Psi &=& (1+\zeta\bar\zeta) \left(
    \frac{1}{1+\bar\zeta^2} \frac{\partial\Psi}{\partial\alpha}
             + \frac{i}{1-\bar\zeta^2} \frac{\partial\Psi}{\partial 
\beta}
            \right) \nonumber \\
   &+& s \zeta \Psi, \\
  \bar\eth \Psi &=& (1+\zeta\bar\zeta) \left(
     \frac{1}{1+\zeta^2} \frac{\partial\Psi}{\partial\alpha}
   - \frac{i}{1-\zeta^2}\frac{\partial\Psi}{\partial\beta}
   \right) \nonumber \\
   &-& s \bar\zeta \Psi,
\label{eq:partial_zeta}
\end{eqnarray}
where the values of $(\zeta, \bar\zeta)$ for each grid point are
computed from the respective $(\xi, \bar\xi)$ values as per Eq.~(\ref
{eq:xi_zeta}).

To complete the prescription of the $\eth$ operator on the cubed sphere we
must specify the transformation rule for spin-weighted functions on the
sphere. The stereographic coordinates on the various patches transform
according to
\begin{subequations}
\begin{eqnarray}
   \displaystyle{\zeta_1 = \frac{\zeta_4-1}{\zeta_4+1}}, &
   \displaystyle{\zeta_2 = \frac{\zeta_1-1}{\zeta_1+1}},
   \nonumber \\
   \displaystyle{\zeta_3 = \frac{\zeta_2-1}{\zeta_2+1}}, &
   \displaystyle{\zeta_4 = \frac{\zeta_3-1}{\zeta_3+1}},
\label{eq:transfeq}
\\
\nonumber \\
   \displaystyle{\zeta_1 =-i\,\frac{\zeta_5+i}{\zeta_5-i}}, &
   \displaystyle{\zeta_2 =-i\,\frac{\zeta_5-1}{\zeta_5+1}},
   \nonumber \\
   \displaystyle{\zeta_3 =-i\,\frac{\zeta_5-i}{\zeta_5+i}}, &
   \displaystyle{\zeta_4 =-i\,\frac{\zeta_5+1}{\zeta_5-1}},
\label{eq:transfneq}
\\
\nonumber \\
   \displaystyle{\zeta_1 = i\,\frac{\zeta_6-i}{\zeta_6+i}}, &
   \displaystyle{\zeta_2 = i\,\frac{\zeta_6-1}{\zeta_6+1}},
   \nonumber \\
   \displaystyle{\zeta_3 = i\,\frac{\zeta_6+i}{\zeta_6-i}}, &
   \displaystyle{\zeta_4 = i\,\frac{\zeta_6+1}{\zeta_6-1}},
\label{eq:transfseq}
\\
\nonumber \\
   \displaystyle{\zeta_6= \frac{1}{\zeta_5}}, &
   \displaystyle{\zeta_3=-\frac{1}{\zeta_1}},
   \nonumber \\
   \displaystyle{\zeta_4=-\frac{1}{\zeta_2}},
\label{eq:transfop}
\end{eqnarray}
\label{eq:transf}
\end{subequations}
\noindent where Eqs.~(\ref{eq:transfeq}) relate neighboring equatorial patches,
Eqs.~(\ref{eq:transfneq}) and (\ref{eq:transfseq}) supply the coordinate
transformations between equatorial patches and the north and south patch,
respectively. Eqs.~(\ref{eq:transfop}), which relate diametrically
opposed patches on the sphere, are not strictly necessary for a
numerical implementation and can be deduced from (\ref{eq:transfneq})
and (\ref{eq:transfseq}). Our convention for the gnomic parameterization
follows Ref.~\cite{ronchi}, and it has the advantage that the orientation
of the $(u,v)$ axes on each patch is chosen so as to reduce the amount
of book keeping needed to transfer information between patches. From
Eq.~(\ref{eq:transf}), adjacent patches with coordinates $\zeta_i$
and $\zeta_j$ are related by $A\zeta_i = (B\zeta_j-1)/(B\zeta_j+1)$,
where $A=1,\pm i$ and $B=1,\pm i$, while opposite patches are related
by $\zeta_i=C/\zeta_j$, with $C=\pm 1$, in particular, we recover
the coordinates transformation between ``north'' ($\zeta_N=\zeta_5$)
and ``south'' ($\zeta_S=\zeta_6$) patches, $\zeta_N=1/\zeta_S$, as in
Ref.~\cite{eth}.

Given the dyad $q^A=P(1,i)$, $P(\zeta,\bar\zeta)=1+\zeta\bar\zeta$,
and expressing $q=q^A\partial_A\equiv P(\zeta,\bar\zeta)
\partial_{\bar\zeta}$ in two coordinate systems $x^A=(\zeta,\bar\zeta)$
and $x^{A'}=(\zeta',\bar\zeta')$, it can be seen that
\begin{equation}
q =  W(\zeta,\bar\zeta) q', \nonumber \\
\end{equation}
where
\begin{equation}
    W(\zeta,\bar\zeta) = \frac{P(\zeta,\bar\zeta)}
                              {P(\zeta',\bar\zeta')}
     \frac{\partial\bar\zeta'}{\partial\bar\zeta} \, ,
\label{eq:Wzzp}
\end{equation}
and where the substitution $\zeta'=\zeta'(\zeta)$ is understood in the
right hand side of~(\ref{eq:Wzzp}). From this we deduce the transformation
rule for spin-$s$ functions on the sphere, given here only for the case
of adjacent patches
\begin{eqnarray}
   \displaystyle{\Psi_1 = \left(\frac{\zeta_1-1}{\bar\zeta_1-1}\right)^s
                              \Psi_4}, &
   \displaystyle{\Psi_2 = \left(\frac{\zeta_2-1}{\bar\zeta_2-1}\right)^s
                              \Psi_1}, \nonumber \\
   \displaystyle{\Psi_3 = \left(\frac{\zeta_3-1}{\bar\zeta_3-1}\right)^s
                              \Psi_2}, &
   \displaystyle{\Psi_4 = \left(\frac{\zeta_4-1}{\bar\zeta_4-1}\right)^s
                              \Psi_3}, \nonumber \\
   \displaystyle{\Psi_1 = \left(-\frac{\zeta_1+i}{\bar\zeta_1-i} 
\right)^s
                              \Psi_5}, &
   \displaystyle{\Psi_2 = \left(i\,\frac{\zeta_2+i}{\bar\zeta_2-i} 
\right)^s
                              \Psi_5}, \nonumber \\
   \displaystyle{\Psi_3 = \left(\frac{\zeta_3+i}{\bar\zeta_3-i}\right)^s
                              \Psi_5}, &
   \displaystyle{\Psi_4 = \left(-i\,\frac{\zeta_4+i}{\bar\zeta_4-i} 
\right)^s
                              \Psi_5}, \nonumber \\
   \displaystyle{\Psi_1 = \left(-\frac{\zeta_1-i}{\bar\zeta_1+i} 
\right)^s
                              \Psi_6}, &
   \displaystyle{\Psi_2 = \left(-i\,\frac{\zeta_2-i}{\bar\zeta_2+i} 
\right)^s
                              \Psi_6}, \nonumber \\
   \displaystyle{\Psi_3 = \left(\frac{\zeta_3-i}{\bar\zeta_3+i}\right)^s
                              \Psi_6}, &
   \displaystyle{\Psi_4 = \left(i\,\frac{\zeta_4-i}{\bar\zeta_4+i} 
\right)^s
                              \Psi_6}.
\label{eq:psitransfeq}
\end{eqnarray}
For the case of functions with spin-weight zero the transformations
reduce to $\Psi_j(\zeta_j,\bar\zeta_j) = \Psi_{i} (\zeta_i,\bar\zeta_i)$.

\section{Numerical implementation}
\label{sec:numerics}

We give here a summary of the numerical techniques used so far in
the LEO framework. It is worth noting that the framework is easily
extensible, and thus we are not restricted, for instance, to the
particular choice of radial grid made here, nor to the choice of radial
or time integration schemes used in the present work. In the subsection
on finite-difference operators on the sphere, for instance, we describe
higher--order extensions that we have elected not to use in the example
application considered here, as they are inconsistent with the radial and
time integration schemes, which we have taken unchanged from~\cite{news}.

\subsection{Radial grid and finite difference operators}

Following~\cite{jcp92}, we take the computational radial grid
to be equally-spaced in the compactified coordinate $x=r/(R+r)$,
restricted to the range $x_B\le x \le 1$, i.e. $x_k=x_B + (k-1)
\Delta x$, $k=1,\ldots,N_x$, $\Delta x = (1-x_B)/(N_x - 1)$, with
$x_B=r_B/(R+r_B)$.  We express radial derivatives in terms of the
compactified grid $x_i$, via the relation $\partial x/\partial r =
(1-x)^2/R$, e.g.
\begin{eqnarray}
    \left.f_{,r}\right|_{k+\frac{1}{2}} &=&
        \frac{(1-x_{k+\frac{1}{2}})^2}{R}
    \frac{\left( f_{k+1} - f_{k} \right)}{\Delta x} ,
\\
    \left.f_{,r}\right|_{k} &=&
        \frac{(1-x_{k})^2}{R}
    \frac{\left( f_{k+1} - f_{k-1} \right)}{2\Delta x}.
    \label{eq:rderivative}
\end{eqnarray}

\subsection{Centered finite difference operators on the sphere}

\begin{figure}[!ht]
\includegraphics[width=3.25in,height=3.25in,angle=0]{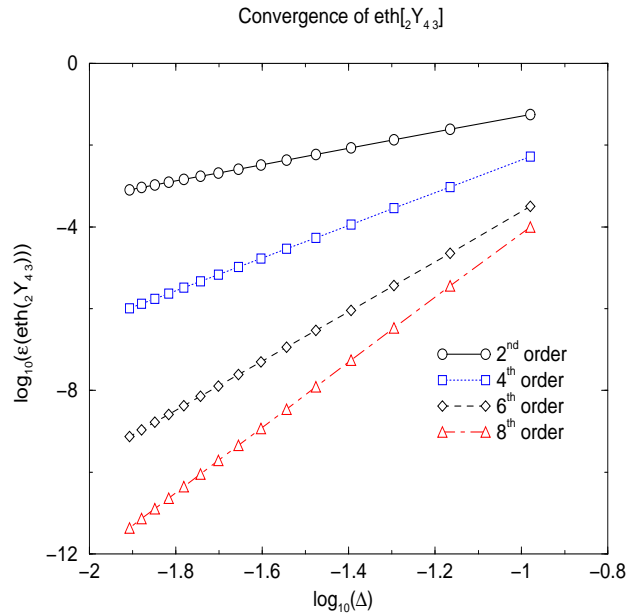}
\caption{Convergence rate of the $\eth$ operator,
built upon angular derivatives of order 2, 4, 6, and 8 (indicated in
the graph as circles, squares, diamonds and triangles, respectively),
acting on ${}_{2} Y_{4\,3}$, and with grid sizes ranging from $N_\zeta=16$
to $N_\zeta=128$.}
\label{fig:eth-conv}
\end{figure}

We construct an equally-spaced grid on the gnomic coordinates
$x^{A}=(\alpha, \beta)$, with $\alpha_i = -\pi/4 + (i-\frac{1}{2})
\Delta$, $\beta_j = -\pi/4 + (j-\frac{1}{2}) \Delta$, and $\Delta =
\pi/(2 N_\xi)$, $i,j = 1\ldots N_\xi$.  The useful part, exclusive of
ghost zones, for each of the coordinates ranges from $-\pi/4+\Delta/2$
to $\pi/4-\Delta/2$. With this arrangement, the points with $|u|=\pi/4$
or $|v|=\pi/4$ are excluded, and thus we avoid storing double values for
the points at the edges of each patch, and triple values for the points on
the corners where three patches meet.  Adding $N_g$ ghost zones on each
side of the grid allows us to evaluate derivatives to order $N=2 N_g$
with centered stencils of the form
\begin{equation}
    \left.\frac{\partial f}{\partial \alpha}\right|_{i,j} =
    \frac{1}{\Delta} \sum_{k=1}^{N/2} c_{k}
    \left( f_{i+k,j} - f_{i-k,j} \right).
    \label{eq:aderivative}
\end{equation}
The coefficients for the derivatives, up to $8$-th order, are given in
Table~\ref{tab:derivatives}.
\begin{table}[!ht]
\begin{ruledtabular}
\begin{tabular}{|l|c|c|c|c|}
$N_g$ & $c_1$  & $c_2$  & $c_3$ & $c_4$ \\
\hline
1  & 1/2   &        &       &        \\
\hline
2  & 8/12  & -1/12  &       &        \\
\hline
3  & 3/4   & -3/20  & 1/60  &        \\
\hline
4  & 4/5   & -1/5   & 4/105 & -1/280 \\
\end{tabular}
\end{ruledtabular}
\caption{Coefficients for centered angular derivatives.}
\label{tab:derivatives}
\end{table}
We can verify that the coefficients of Table~\ref{tab:derivatives}
for each of the derivatives are correct by noting that the numerical
error of derivatives of order $N$ is within the level of round-off when
applied to a polynomial test function $F$ of order $N$ or lower. We
have also verified the proper convergence rate of the numerical $\eth$
and $\bar\eth$ operators when applied to spin-weighted spherical
harmonics~\cite{mode}.  Fig.~\ref{fig:eth-conv} shows the proper
convergence rates of the $\eth$ operators constructed from derivatives
of second, fourth, sixth, and eighth order when applied to the spin-2
spherical harmonic ${}_2 Y_{4\,3}$ on grid sizes ranging from $N_\zeta=16$
to $N_\zeta=128$.

\subsection{One-dimensional interpolation of ghost zones}

\begin{figure}[!ht]
\includegraphics[width=3.25in,height=3.25in,angle=0]{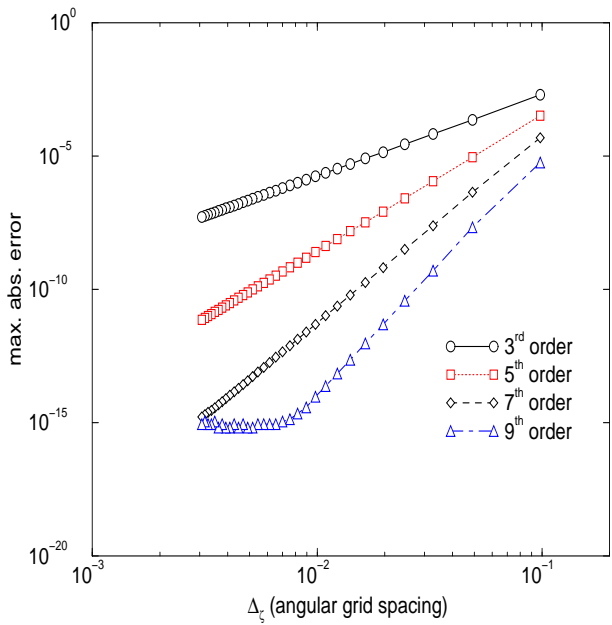}
\caption{Convergence rate of the various interpolation schemes used.
Shown in the graph are the $3^{rd}$ order (circles), $5^{th}$ order (squares), 
$7^{th}$ order (diamonds), and $9^{th}$ order (triangles) interpolators.
For the highest order interpolator used ($9$-th order), the error goes
down to double-precision round-off level ($\sim 10^{-16}$) when more
than $N_\zeta=100$ angular points per patch are used.}
\label{fig:interp}
\end{figure}

Since the ghost zones required to evaluate derivatives fall on coordinate
lines parallel to the boundary, we can obtain function values at these
ghost zones with one--dimensional interpolations. We use standard
Lagrangian interpolation formulae to $N$-th order accuracy,
\begin{equation}
    f(x) = \sum_{i=1}^{N} f_i
    \prod_{j \ne i} \frac{(x - x_j)}{(x_i - x_j)} ,
    \label{eq:interpolator}
\end{equation}
adapted to equally spaced grids, i.e. $x_j=x_0+j\,\Delta$.
Fig.~\ref{fig:interp} shows the calibration of the interpolation routines
with a test function consisting of a polynomial of order $15$, i.e.
\begin{equation}
    P_{N}(\alpha)=\sum_{i=0}^{N} c_{i}\, \alpha^{i}, \quad
   -\frac{\pi}{4} \le \alpha \le \frac{\pi}{4}
\end{equation}
for $N=15$, where the coefficients $c_{i}, i=1\ldots N$ are chosen
randomly, subject to the condition $|c_{i}| \le 1$. The interpolants
display convergence to the correct order (3, 5, 7, and 9-th order,
respectively), for grid sizes in the range $8 \le N_\zeta \le 256$.
For this range of values ($8 \le N_\zeta \le 256$), there are from 32
to 1024 points in the great circles determined by the intersection of
the sphere with any of the Cartesian coordinate planes. As expected,
for smooth data such as our test function, for sufficiently large grid
sizes, the error goes down to round-off level when using the higher-order
schemes. This saturation effect is already visible in the plot for the
$9$-th order interpolator when angular grid sizes reach approximately
$N_\zeta=100$. As indicated in Sec.~\ref{sec:gnomic}, we have chosen
the range of the spherical coordinates $(\alpha,\beta)$ so that there
are no overlapping points at the edge of each patch. This avoids the
awkward procedure of averaging values from different patches to obtain
a single-valued function throughout the computational grid.

The number of ghost zones, the order of the finite difference
approximations and the order of interpolation, while related, are not
directly tied to each other. One requirement is that we must have enough
ghost zones ($N_g$) to compute the finite-difference approximation to
the desired order, $N_F$; and since in general we want to use centered
differences, the relation $N_F \le 2 N_g$ must hold. If we wish to
maintain the symmetry of the interpolation stencils, $N_I \le 2 N_g +
1$ must also hold. For the cases we have considered, we find that our
algorithms are stable if $N_I \ge N_F + 1$, with the inequality required
only in the case of $N_F=2$, the lowest order of finite-differences that
we considered. We are otherwise free to vary the number of ghost zones
as dictated by efficiency considerations.

\subsection{Integrals over the sphere and volume integrals}

\begin{figure}[!ht]
\includegraphics[width=3.25in,height=3.25in,angle=0]{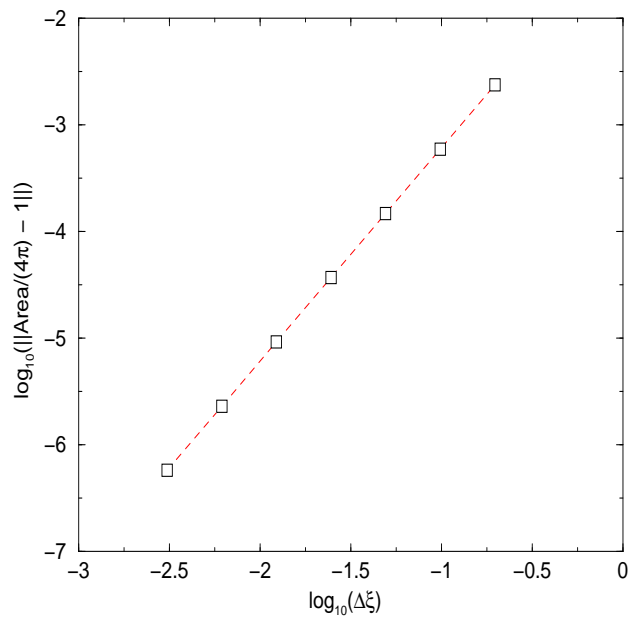}
\caption{Convergence rate of the integral of the area element over the
sphere, for grid sizes ranging from $N_\zeta=8$ to $N_\zeta=512$. The markers
indicate the error of the area element at the corresponding resolution, the line is the least-squares fit, yielding a convergence rate of $2.0$.}
\label{fig:integral-conv}
\end{figure}

Integrals over the sphere and volume integrals arise naturally, in
particular when computing norms of various quantities. We evaluate
integrals on the sphere to second order accuracy by evaluating the area
element in gnomic coordinates,
\begin{equation}
    d \Omega = \frac{(1 + U^2) (1 + V^2)}
                    {(1 + U^2 + V^2)^{3/2}}
               \Delta\alpha \Delta\beta,
\label{eq:area}
\end{equation}
evaluating the function value on grid cell centers, $f_{i,j} =
f(U_i,V_j)$, and summing over grid cells,
\begin{equation}
    \int_{S} f\, d\Omega = \sum_{i=1}^{N_\zeta} \sum_{j=1}^{N_\zeta}
                   f_{i,j}
                   \frac{(1 + U_i^2) (1 + V_j^2)}
                        {(1 + U_i^2 + V_j^2)^{3/2}} \Delta^2,
    \label{eq:sph_int}
\end{equation}
where $\Delta$ stands for the grid spacing on both coordinates
($\alpha,\beta)$, which we have taken to be the same. Note that
since the spherical patches do not overlap, the integral over the
sphere is just the sum of the integrals over the individual patches.
Fig.~\ref{fig:integral-conv} shows the converge of the integral of
the area element itself to the correct answer of $\int d\Omega = 4\pi$
for grid sizes in the range of $N_\zeta=8$ to $N_\zeta=512$.
The measured convergence rate is $2.0$, in full agreement with the
expected result.

Volume integrals are computed similarly to integrals over the sphere,
but in this case evaluating the spherical contributions mid-point in
between radial points, i.e.
\begin{eqnarray}
    \int_{S} 
     d\Omega
    \int_{r_m}^{r_n} 
     f 
     r^2 \,dr \,
        =  
           \sum_{i=1}^{N_\zeta} 
           \sum_{j=1}^{N_\zeta}
                   \frac{(1 + U_i^2) (1 + V_j^2)}
                        {(1 + U_i^2 + V_j^2)^{3/2}} 
                   \Delta^2 
                   \nonumber \\
   \times   \sum_{k=m}^{n-1} 
                   \frac{x_{k+\frac{1}{2}}^2}
                        {(1 - x_{k+\frac{1}{2}})^4} 
                   \frac{ \left(f_{i,j,k} + f_{i,j,k+1}\right)}{2}
                   \,\Delta x\,  .
    \label{eq:vol_int}
\end{eqnarray}
We replace the flat volume element, $dV=r^2dr\,d\Omega$, with the volume
element corresponding to a Bondi metric, $dV=r^2 e^{2\beta}dr\,d\Omega$,
when appropriate.  To speed up the evaluation of integrals, we pre-compute
the area element on the sphere, Eq.~(\ref{eq:area}).

\subsection{Accuracy of the spin-weighted spherical harmonic decomposition}

\begin{figure}[!ht]
\includegraphics[width=3.25in,height=3.25in,angle=0]{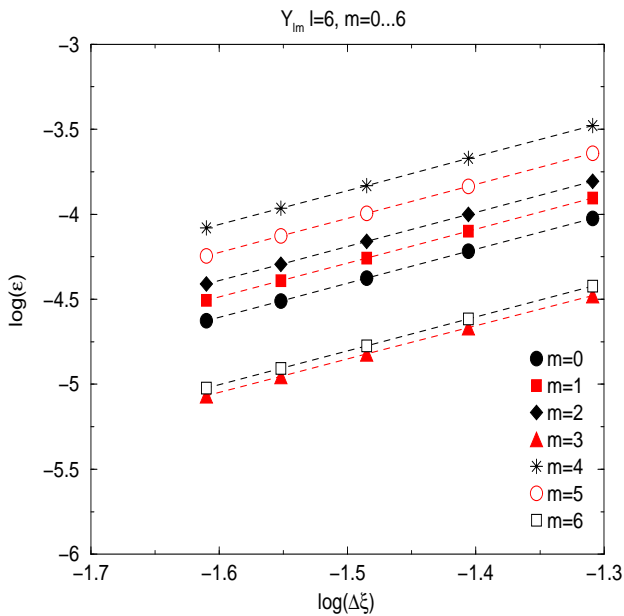}
\caption{Convergence of the orthonormality condition, illustrated here
by computing the convergence rate of $\int_{S} {}_{0} Y_{l\,m} {}_{0}
\bar{Y}_{l\,m} \equiv 1$, for the case $l=6$, $m=0\ldots 6$, on grid
sizes ranging from $N_\zeta=32$ to $N_\zeta=64$}
\label{fig:F_l=6}
\end{figure}

\begin{figure}[!ht]
\includegraphics[width=3.25in,height=3.25in,angle=0]{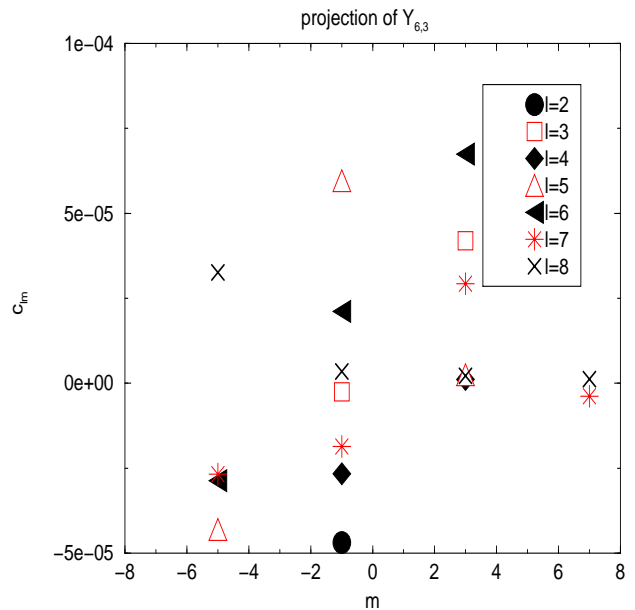}
\caption{Error in the coefficients $c_{l\,m}$ when the function being
projected is the spherical harmonic $Y_{6\,3}$, on a grid of $N_ \zeta=64$
points. Coefficients whose error is at round--off level are not shown.}
\label{fig:proj_error}
\end{figure}

We make use of spin-weighted spherical harmonics ${}_{s} Y_{lm}$
throughout this paper, following the convention of~\cite{mode}.  In order
to estimate the error introduced when we perform a spin-weighted
spherical harmonic decomposition, we look at how well the orthonormality
condition
\begin{equation}
    \int_{S} {}_{s} Y_{lm} \ {}_{s} \bar{Y}_{lm} d\Omega =
    \delta_{l,l'} \delta_{m,m'},
\label{ylm_ortho}
\end{equation}
is preserved (at the numerical level) for spherical harmonics with
spin-weight $s=0$, 1 and 2, for a range of values of $\ell$ and $m$
and angular grid sizes. As expected, the numerical value of the
integral converges to the analytic result to second order on the
grid spacing, since we have chosen to use a second-order integration
algorithm. Fig.~\ref{fig:F_l=6} illustrates one instance, where we have
taken $s=0$, $\ell=6$, with $m=-6\ldots 6$, and varied the angular grid
size from $N_ \zeta=32$ to $N_\zeta=64$.
We can also place an estimate on the accuracy of the projection
of a spin--weight $s$ function,
\begin{equation}
    c_{lm} [F] = \int_{S} F \ {}_{s} \bar{Y}_{lm} d\Omega ,
\end{equation}
based on the magnitude of the off-diagonal values in~(\ref{ylm_ortho})
for a given grid size. When projecting the test functions $Y_{l'\,m'}$
into the spherical harmonics $Y_{l\,m} $ for $l=0\ldots l_{max}$,
$m=-l\ldots l$, at the analytic level we would expect to obtain zero for
all coefficients, except for $c_{l'\,m'}$ which would be identically one.
We find that grid sizes of $N_\zeta=64$ and larger are sufficient to
keep the error in the coefficients to within one part in $10^{4}$, which
again is consistent with our integration scheme being second-order in
the angular discretization. Fig.~\ref{fig:proj_error} shows the error in
the coefficients computed for $Y_{l'\,m'}$, $l'=6$, $m'=3$ on a grid with
$N_\zeta=64$ points. We have omitted from the graph those coefficients
for which the error is already at the level of round-off.

The preceding description of the numerical implementation is complete
but for one key aspect, namely our parallelization strategy. In our
framework, the six cubed-sphere grid patches are decomposed into {\it
computational} sub-patches, each with the same number of points on
the angular directions, for efficiency reasons. These sub-patches
are distributed among processors, and the ghost-zone values required
for the computation of angular derivatives are communicated by the
use of message-passing calls~\cite{gropp}. The radial direction is not
distributed, as the characteristic algorithm requires a radial march for
the integration of the hypersurface equations as well as the evolution
equations~\cite{jcp92}.  The computational complexity of a parallel
implementation via message passing lies in that, knowing the location
of its assigned grid sub-patch on the global grid, each processor must
determine which processors are its nearest neighbors, i.e. to which
processes it must supply ghost-zone information (and also receive that
information from).  Due to the relative orientation of the cubed-sphere
patches, we need to know whether the order in which the ghost zones are
traversed must be reversed for sub-patches on the edge of a cubed-sphere
patch. Since the sub-patch to processor mapping remains constant during
a simulation, the relevant information needs to be computed only once,
and at any rate, it incurs no measurable overhead in the computation
involved in a simulation. An efficient and scaling implementation of
the message passing itself requires only a small subset of the full
MPI functionality:  a few calls to set up the appropriate groups of
processors; sends, receives and waits (for ghost zone communication);
some additional reduction operations (to accumulate integrated values),
and some broadcasts (to propagate parameters). Exclusive of file access
operations, only 14 MPI functions in all are invoked.

Having established that all the key computational aspects of the framework
are in place, and have been correctly implemented, we proceed next to
consider specific applications of the framework to systems of physical
interest.

\section{A three--dimensional massless scalar field scattered off a
Schwarzschild black hole} \label{sec:ekg3d}

We use the numerical formalism developed in the preceding sections to
solve numerically a model problem consisting of a self-gravitating
massless scalar field in three dimensions.  Our starting points
are Ref.~\cite{reduced} for a description of the vacuum problem,
and Ref.~\cite{bsglrw05} for the coupling of the scalar field to
the gravitational metric fields. We use coordinates based upon a
family of outgoing null hypersurfaces, and we let $u$ label these
hypersurfaces, $x^A$ ($A=2,3$) label the null rays and $r$ be a surface
area coordinate. In the resulting $x^\alpha=(u,r,x^A)$ coordinates,
the metric takes the Bondi--Sachs form \cite{bvm62,s62}
\begin{eqnarray}
ds^{2}&=&-[e^{2\beta }(1+W/r)-r^{2}h_{AB}U^{A}U^{B}]du^{2}-
2e^{2\beta }du dr \nonumber \\
&-&2r^{2}h_{AB}U^{B}dudx^{A}+r^{2}h_{AB}dx^{A}dx^{B},
\end{eqnarray}
where $W$ is related to the more usual Bondi--Sachs variable
$V$ by $V=r+W$, and where $h^{AB}h_{BC}=\delta ^{A}_{C}$  and
$det(h_{AB})=det(q_{AB})$, with $q_{AB}$ a unit sphere metric, given
in terms of a complex dyad $q_A$ satisfying $q^A q_A = 0$, $q^A
\bar q_A = 2$, $q^A=q^{AB}q_B$, with $q^{AB} q_{BC} = \delta^A_C$
and $q_{AB}=\frac{1}{2}(q_A \bar q_B + \bar q_A q_B)$.  We also use
the intermediate variable $Q_{A}=r^{2}e^{-2\beta }h_{AB}U^{B}_{,r}$.
We represent tensors on the sphere by spin-wighted variables \cite{eth}.
The conformal metric $h_{AB}$, is represented by the complex function
$J=h_{AB}q^{A}q^{B}/2$, and by the real function $K=h_{AB}q^{A}\bar
q^{B}/2$, where $K^{2}=1+J\bar J$. The metric functions $U^A$ are
similarly encoded in the complex function $U=U^{A}q_{A}$.  Thus,
it is necessary to introduce the intermediate spin-weighted variable
$Q=Q_{A}q^{A}$, as well as the (complex differential) operators $\eth$
and $\bar\eth$ (see \cite{eth} for full details).

Treating the Einstein-Klein-Gordon model problem consistently within
the LEO framework requires some modifications to~\cite{bsglrw05},
specifically to the wave equation for the scalar field ($ \square \phi =0$)
which is given by Eqs.~(21)-(27) of~\cite{bsglrw05}.  We substitute all
second-order angular derivatives of the metric fields in terms of $\eth$
and $\bar\eth$ operators acting on the additional fields $\nu=\bar\eth
J$, $k=\eth K$ and $B=\eth\beta$ introduced in Ref.~\cite{reduced},
whenever possible. A consistent treatment is obtained by introducing
the additional variable
\begin{equation}
   \psi=\eth\chi, \label{eq:psi_def}
\end{equation}
where $\chi=r\phi$, so that the scalar field equation is also in first-order 
differential form in the angular variables, on a par with the
approach of~\cite{reduced} for the metric equations. The Bondi--Sachs
hierarchy of hypersurface equations,
\begin{eqnarray}
   \nu_{,r} &=& \bar\eth J_{,r}\, ,
   \label{eq:nu} \\
   k_{,r} &=& \eth K_{,r}\, ,
   \label{eq:k} \\
    \beta_{,r} &=& \frac{r}{8} \left( J_{,r} \bar J_{,r} - K^2_{,r}\right)
              +2\pi r(\phi_{,r})^{2},
   \label{eq:beta} \\
   B_{,r} &=& \eth\beta_{,r}\, ,
   \label{eq:B} \\
     (r^2 Q)_{,r}  &=&
      r^2 \bigg[ - K ( k_{,r} + \nu_{,r} )
                 + \bar \nu J_{,r} + \bar J \eth J_{,r}
                 + \nu K_{,r} \nonumber \\
                 && + J \bar k_{,r} - J_{,r} \bar k
           \bigg]  \nonumber \\
   & + &  \frac{r^2}{2K^2}
          \left[ \bar \nu \left(J_{,r} - J^2 \bar J_{,r} \right)
                + \eth J \left(\bar{J}_{,r} - \bar{J}^2 J_{,r} \right)
          \right] \nonumber \\
   &&  +  2 r^2 B_{,r} -4r B
                  +16\pi r\phi _{,r}\psi\, ,
     \label{eq:Q} \\
   r^2 U_{,r}  &=& e^{2\beta} \left( K Q - J \bar Q \right),
     \label{eq:U}
\\
     (r^2 \tilde{W})_{,r}  &=& \Re \Bigg\{ e^{2\beta}
    \left( \frac{\cal R}{2} -K \left( \bar\eth B + B \bar B\right)
   + \bar J \left( \eth B + B^2 \right) \right.
\nonumber \\
& + & \left. (\nu - k) \bar B \right)
   -   1 + 2\,r \bar \eth U + \frac{r^2}{2} \bar\eth U_{,r}
\nonumber \\
&&   -  e^{-2 \beta} \frac{r^4}{4}
 \bar U_{,r} \left( K U_{,r} + J \bar U_{,r} \right) \Bigg\}
\nonumber \\
  & - & 2\pi \frac{e^{2\beta }}{r^2} \left[
        2K\bar{\psi} \psi -J \bar{\psi}^{2}
        -\bar{J} \psi^{2}
                         \right]\, ,
 \label{eq:W}
\\
   \psi_{,r} &=& \eth\chi_{,r}\, ,
   \label{eq:psi}
\end{eqnarray}
now includes an additional consistency condition, Eq.~(\ref{eq:psi}),
and the equations for $\beta$, $Q$ and $\tilde W=W/r^2$ are modified to
include the source terms as shown above. The evolution equation for the
metric field $J$ is given by
\begin{eqnarray}
   &&2 \left(rJ\right)_{,ur}
    - \left(r^{-1}V\left(rJ\right)_{,r}\right)_{,r} =
     - K \left( r \eth U_{,r} + 2\, \eth U \right)
 \nonumber \\
    &&+ \frac{2}{r} e^{2\beta} \left( \eth B + B^2 \right)
    - \left(r \tilde{W}_{,r} + \tilde{W} \right) J
      + J_{H} + J P_u
\nonumber \\
   &&+\frac{8\pi}{r^3}e^{2\beta }\psi^{2} ,
    \label{eq:J}
\end{eqnarray}
with the quantities ${\cal R}$, $J_H$ and $P_u$ as in Eqs.~(24)--(26)
of~\cite{reduced}.  The scalar field evolution equation follows from
Eq.~(21) of~\cite{bsglrw05},
\begin{equation}
   2 \chi_{,ur}-\left(\frac{V}{r}  \chi_{,r}\right)_{,r} = 
   -\left(\frac{W}{r}\right)_{,r}\frac{\chi}{r} +N_{\phi} .
\label{eq:phi}
\end{equation}
The source term $N_{\phi}$ is
\begin{eqnarray}
    N_{\phi}&=&
   \frac{e^{2\beta}}{r} \Big[
      - \frac{1}{2r} \left( \bar{J} \eth \psi + J \bar{\eth}\bar{\psi} \right)
      + \frac{K}{r} \bar{\eth} \psi
   \nonumber \\
  &&  + \left( K \bar{B} -\bar{J} B - \frac{1}{2} (K \bar{Q} - \bar{J} Q)
               - \frac{\bar{\nu}}{2}  \right. \nonumber \\
  &&
        \left. + \frac{1}{4K} ( \bar{J} \nu + J \bar{\mu} )
        \right) \frac{\psi}{r}
        \nonumber \\
  &&  + \left( K B - J \bar{B} - \frac{1}{2} (K Q - J \bar{Q})
               - \frac{     \nu }{2}  \right. \nonumber \\
  &&
        \left. + \frac{1}{4K} ( J \bar{\nu} + \bar{J} \mu )
        \right)\frac{ \bar{\psi}}{r}
                        \Big]
        \nonumber \\
  &&
  - \frac{1}{r}(U \bar{\psi} + \bar{U} \psi)
  - \frac{r}{2} \phi_{,r} ( \bar{\eth} U + \eth\bar{U} )
        \nonumber \\
  &&
  - [ U (\bar{\psi}_{,r}-\bar\psi) + \bar{U}( \psi_{,r}-\psi )].
\label{eq:Nphi}
\end{eqnarray}
Following~\cite{quasi}, we have used the shorthand $\mu=\eth J$,
and eliminated the radial derivatives $U_{,r}$ and $\bar{U}_{,r}$ using
Eq.~(\ref{eq:U}),
\begin{equation}
   Q = r^2 e^{-2\beta} (K U_{,r} + J \bar{U}_{,r}) .
\end{equation}

The data required on the initial null cone are the evolution variables
$J$ and $\phi$.  Given boundary values at a fixed value of $r$, the
remaining variables ($\nu$, $k$, $\beta$, $B$, $Q$, $U$ and $\tilde{W}$)
can be determined on the initial null cone by explicit integration of the
hypersurface equations (see \cite{bsglrw05} for details). The evolution
equations (\ref{eq:J}) and (\ref{eq:phi}) can then be used to find $J$
and $\phi$ on the next null cone, and the process repeated to determine
the spacetime to the future of the initial slice.

\subsection{Scalar field on a fixed background}

The above system of equations describes a self--gravitating scalar
field. In the limit of small amplitudes, $|\phi| << 1$, the scalar field
can be treated as a perturbation propagating on a fixed background.
This considerably simpler model is contained in the fully nonlinear case,
and is implemented in our code by integrating only Eqs.~(\ref{eq:psi})
and ~(\ref{eq:phi}). For a Scharwzschild background, the metric fields
$J$, $\beta$, $U$, $\nu$, $k$ and $B$ are zero, and $V=r-2M$.  The source
term in Eq.~(\ref{eq:Nphi}) reduces to $N_\phi=\bar\eth\psi/r$, and we
are left with the system
\begin{eqnarray}
 2 \chi_{,ur}-\left( \left(1 - \frac{2M}{r} \right) \chi_{,r} \right)_{,r} &=& 
   -\frac{2 M \chi}{r^3}
   + \frac{\bar\eth\psi}{r} \ ,
\nonumber \\
   \psi_{,r} &=& \eth\chi_{,r}\ ,
  \label{eq:l_phi}
\end{eqnarray}
For the simulations we discuss in the present work, we will be interested
in solutions of the scalar field on a fixed background with definite
angular dependence, as discussed in the next sub-section.

\subsection{Quasi-normal modes in a Schwarszchild background}

The linear equation for the scalar field on a fixed background,
Eq.~(\ref{eq:l_phi}) is separable, i.e. its solutions can be written in
the form
\begin{equation}
\phi(u,r,x^A) = \sum_{\ell=0}^{\infty} \sum_{m=-\ell}^{\ell} 
       \chi_{\ell\,m}(u,r)  \frac{Y_{\ell\,m}(x^A)}{r}\, ,
\end{equation}
with the $x^A$ coordinates in the sphere, and where each of the
$\chi_{\ell\,m}$ satisfies the one-dimensional wave equation in the
plane $(u,r$)
\begin{eqnarray}
2\,\chi_{,ur} - \left( \left(1-\frac{2M}{r} \right) \chi_{,r}\right)_{,r} = 
\nonumber \\
- \left(\frac{2M}{r^3}+ \frac{\ell(\ell+1)}{r^2} \right) \chi.
\label{eq:sph}
\end{eqnarray}
where we have used the property $\eth\bar\eth \chi = -\ell(\ell+1)\chi$
\cite{np66}.  Eq.~(\ref{eq:sph}) is the usual equation governing the
scalar perturbations of a Schwarzschild black hole~\cite{nollert},
written here in characteristic coordinates $(u,r,x^A)$.  It can be put
in a more familiar form by writing it in the coordinates $(t,r_*)$,
with $u=t-r_*$, and where $r_*$ is the usual ``tortoise'' coordinate,
$r_*=r+2M \ln(r/2M-1)$.
\begin{equation}
\chi_{,tt} - \chi_{,r_*r_*} + \hat{V}(r) \chi = 0 ,
\label{eq:qnm}
\end{equation}
where $\chi = r \phi$ and the potential $\hat{V}(r)$ is given by
\begin{equation}
\hat{V}(r) = \left( 1 - \frac{2M}{r} \right) 
  \left[\frac{2M}{r^3}+ \frac{\ell(\ell+1)}{r^2} \right] \chi ,
\label{eq:potential}
\end{equation}
and we have denoted it by $\hat{V}$ to avoid confusion with Bondi's $V$
which we use throughout this paper. Eq.~(\ref{eq:qnm}) has been studied
extensively~\cite{nollert,nollert-schmidt,konoplya}, its most salient
feature being the existence of quasi-normal modes, whose frequencies have
been tabulated, see for example~\cite{konoplya}. Note that the right-hand
side of Eq.~(\ref{eq:sph}) is the correct form of the potential in
$(u,r)$ coordinates. It differs by a factor of $(1-2M/r)$ from the
potential as given in Eq.~(\ref{eq:qnm}), see ~\cite{nollert}, because
that factor is precisely the Jacobian of the coordinate transformation,
$\partial_r/\partial_{r_*}=1-2M/r$.

In the remainder of the present work we will use both the
quasi-normal mode equation, Eq.~(\ref{eq:sph}), and the linear system,
Eq.~(\ref{eq:l_phi}), as tests of the validity of our numerical
implementation. We do this in an incremental fashion, solving
Eq.~(\ref{eq:sph}) for fixed values of $\ell$, and comparing the
effectiveness of the numerical integration scheme and of our boundary
conditions in reproducing the quasi-normal modes. To this end, we
implement a purely radial code for Eq.~(\ref{eq:sph}) that employs the
same numerical integration scheme that is used in the ``linear'' code
(which solves Eq.~(\ref{eq:sph}) and Eq.~(\ref{eq:l_phi})), and in the full
nonlinear code. Using this radial code, we can isolate the effects
arising from the inner-boundary treatment at $r=2M$ by implementing
Eq.~(\ref{eq:sph}) as indicated, in outgoing null coordinates, using
both a non-compactified coordinate $r$, with a simple extrapolative
boundary condition at the outer boundary $r_{out} > 2 M$, and the
compactified coordinate $x=r/(r+R)$, where the outer boundary lies at
future null infinity. The use of a non--compactified coordinate allows
us to isolate any effects that may arise due to the non-uniform
coordinate velocity introduced by the compactified coordinate
$x$. Conversely, simulations using the compactified coordinate avoid
the effects of placing the outer boundary at a finite distance.

We also implement the equivalent of Eq.~(\ref{eq:sph}) in {\it ingoing}
null coordinates, $(v,r)$, with $v=t+r_*$, namely
\begin{eqnarray}
2\, \chi_{,vr} + \left[(1-\frac{2M}{r})\chi_{,r}\right]_{,r} = 
\nonumber \\
\left[\frac{2M}{r^3}+ \frac{\ell(\ell+1)}{r^2} \right] \chi.
\label{eq:sph_v}
\end{eqnarray}
In ingoing null coordinates, the slices at $v=const$ penetrate the
event horizon $r=2M$, effectively providing for an excision scheme,
where evolution can be stopped at a finite number of points inside the
boundary, because the behavior of the field inside the horizon does
not affect the solution outside. Evolutions in ingoing coordinates
are carried out on a non-compactified radial grid, for which boundary
data are required at a fixed value of $r_{out} > 2M$. Because of the
presence of this outer boundary, simulations in ingoing coordinates
can only be run for a limited time, typically $u \sim 2\, r_{out}$,
before outer boundary effects influence the signal extracted. A
similar effect is seen when using outgoing, non--compactified null
coordinates. When using compactified coordinates, no such effects are
seen, as expected. A detailed comparison between ingoing and outgoing
versions of characteristic systems of equations, in compactified as well
as non--compactified coordinates, along with their relative advantages
and disadvantages for specific applications, is worthwhile but lies outside the scope of the
present work and will be reported elsewhere. We will refer only briefly
to these issues in the remainder of this work.

\subsection{Energy carried out by the scalar field}

As a useful physical indicator we calculate the balance of the
scalar field energy contained between the inner boundary and null
infinity. The expressions we give here are valid in the linear case,
where the background metric is that of Schwarzschild. For a more general
approach to this issue, the linkage integrals have to be calculated,
specifically the asymptotic Killing vector field must be parallely
propagated from null infinity~\cite{wt65}.

Restricted to the background case then, given a Killing vector field
$\xi^{\nu}$ of the metric $g_{\mu\nu}$, $\pounds_\xi g_{\mu\nu} = 0$,
we can define the conserved quantity
\begin{equation}
   \mathcal{C} = \int T^{\mu}_{\nu} \xi^{\nu} d\Sigma_{\mu}.
\end{equation}
In particular, selecting the time--like Killing vector
$\xi^{\nu}=\delta^{\nu}{}_{u}$, and for a surface of constant $u$,
$\mathcal{C}$ is the energy contained on the surface,
\begin{equation}
   E(u) = \int T^{u}_{u} dV,
\label{eq:energy}
\end{equation}
where $dV$ is the volume element of the surface at constant $u$.  For a
sphere at constant $r$, $\mathcal{C}$ represents the energy flux across
the surface,
\begin{equation}
   P(u) = \int T^{r}{}_{u} r^2 d\Omega,
\label{eq:flux}
\end{equation}
with $d\Omega$ the solid angle element. The relevant components of the
stress--energy tensor for a massless scalar field are
\begin{eqnarray}
  T^{u}{}_{u} &=&  e^{-2\beta} \frac{V}{2r} \left(\phi_{,r} \right)^2
             + \frac{K}{2\,r^2} \eth\phi \bar\eth\phi
     \nonumber \\
            &-& \frac{1}{4\,r^2} \left[
                                    \bar{J} \left( \eth\phi \right)^2
                                   + J \left( \bar\eth\phi \right)^2
                             \right] \nonumber \\
             &-& \frac{1}{2} e^{-2\beta} \phi_{,r} \left(
                    \bar U \eth\phi + U \bar\eth\phi
                                                \right)\, ,
  \\
  T^{r}{}_{u} &=&  e^{-2\beta} \phi_{,u} \nonumber \\
           &\times&\left[
                   \phi_{,u} -\frac{V}{r} \phi_{,r}
                 + \frac{1}{2} \left( \bar U \eth\phi + U \bar\eth\phi \right)
                              \right] \ .
\end{eqnarray}
In the case of a linear scalar perturbation on a Schwarzschild background,
the energy content of a hypersurface at constant $u$ is given by
\begin{equation}
E(u)=\frac{1}{2}\int \left[ \left(1-\frac{2M}{r} \right) 
                            \left(r\phi_{,r}\right)^2 
   + \eth\phi \bar\eth\phi \right]dr d\Omega \ .
\label{eq:econs}
\end{equation}
The power radiated at time $u$ across a surface of constant $r$, such
as the  inner boundary, which in our simulations we place close enough
to the Schwarzschild black hole, is
\begin{equation}
P_{in}(u)= \int \phi_{,u} \left[ \phi_{,u} -\left(1-\frac{2M}{r}\right) 
\phi_{,r}\right]r^2d\Omega \ .
\end{equation}
For the flux across the inner boundary, the integral as well as the
spatial and time derivatives are to be taken as evaluated at $r=r_{in}$.
Analogously, the  power radiated at time $u$ at null infinity is the
limiting form (as $r\rightarrow\infty$) of the above expression, i.e.
\begin{equation}
P_{out}(u) = \int (r\phi_{,u})^2 d\Omega\, ,
\end{equation}
where we have used the behavior of the scalar field near null infinity
$\mathscr{I}$ to simplify the expression.  With these definitions,
the following global energy conservation law holds
\begin{equation}
\Sigma(u)=E(u) + \int^{u}_{u_0} [P_{out}(u') - P_{in}(u')] du' \equiv const.
\label{eq:gecons}
\end{equation}

Even though the expressions given above hold {\it only} in the limit
in which $\partial_t$ is a Killing vector of the metric, we expect
them to hold in an approximate sense for our nonlinear evolutions, so
we use them as a criterion for code testing.

As stated previously, we use the radial code to
calibrate the fully nonlinear, three-dimensional LEO code in the
linear regime. When computing the energy in the radial code, we
make use of the property
\begin{equation}
 \int \eth\phi \, \bar\eth\phi \, d\Omega=-\int\phi \, 
 \eth\bar\eth\phi \, d\Omega.
\label{eq:angint}
\end{equation}
(see \cite{np66}). Since the data we pose
are pure spherical harmonics, the integral in the right--hand
side is proportional to the norm $\int\phi \bar\phi \,
d\Omega$. Eq.~(\ref{eq:angint}) allows us then to properly account for
the contribution of the angular derivatives of the field to the energy
(\ref{eq:energy}) when using only the radial code.

\section{Additional numerical considerations}
\label{sec:addnumer}

\subsection{ Hypersurface equations}

The integration of the hypersurface equations does not present any
inherent difficulty as they are discretized at mid-point between grid
points as per~\cite{bsglrw05,reduced}. An important issue which arises
because of the parallel implementation of our algorithm is that after each
step in the radial march, that is, after each hypersurface equation has
been advanced radially one grid point, we must synchronize the variable
which has just been integrated. By this we mean that we communicate
the ghost zone values to the processors carrying out the integration in
neighboring patches. Since communication is an expensive operation even
on the most tightly coupled parallel computers, we take the approach of
explicitly synchronizing a variable only if an $\eth$ (or $\bar\eth$)
operator will be applied to the variable in question. An alternative
approach would be to incorporate the synchronization into the $\eth$ (and
$\bar\eth$) operators. The first approach requires more book-keeping
on our part, whereas the second is more straightforward. Because of
the number of $\eth$ (or $\bar\eth$) operations that appear in the full
nonlinear equations, however, the performance difference between these two
approaches is significant. For this reason we take the first approach,
reducing to the minimum possible the amount of communications, with a
substantial increase in performance.

\subsection{Evolution equations}

The evolution equation (\ref{eq:J}) for $J$ is treated as reported in
\cite{bsglrw05}, except that the first two radial points are subject to
the boundary condition explained below.  The evolution equation for the
scalar field is recast in terms of the two--dimensional wave operator
\begin{equation}
  \square^{(2)}\chi=e^{-2\beta}[2\chi_{,ru}-(r^{-1}V\chi_{,r})_{,r}],
  \label{eq:swe}
\end{equation}
where $\chi=r\phi$ and Eq.~(\ref{eq:phi}) reduces then to
\begin{equation}
e^{2\beta}\square^{(2)}\chi={\cal H}, \label{Phi}
\end{equation}
where
\begin{equation}
{\cal H}=-(W/r)_{,r}\chi/r + N_{\phi}.
\end{equation}
Since all two--dimensional wave operators are conformally flat, with
conformal--weight $-2$, we can apply to (\ref{Phi}) a flat--space identity
relating the values of $\chi$ at the four corners $P$, $Q$, $R$ and $S$
of a null parallelogram ${\cal A}$, with sides formed by incoming and
outgoing radial characteristics. In terms of $\chi$, this relation leads
to an integral form of the evolution equation for the scalar field
\begin{equation}
\chi_Q=\chi_P+\chi_S-\chi_R+\frac{1}{2}\int_{\cal A} du\,dr{\cal H}.
\label{rphi_Q}
\end{equation}
The corners of the null parallelogram cannot be chosen to lie exactly
on radial grid points, thus the values of $\chi$ at the vertices of
the parallelogram are approximated to second--order accuracy by linear
interpolations between nearest neighbor--grid points on the same outgoing
characteristic. Approximating the integrand by its value at the center $C$
of the parallelogram (evaluated using average values between the points
$P$ and $S$), we have then
\begin{eqnarray}
\chi_Q &=& \chi_P+\chi_S-\chi_R\nonumber \\
&+&\frac{\Delta u}{4} \left(r_Q-r_P+r_S-r_R \right){\cal H}_C.
\end{eqnarray} 
The evolution algorithm for the metric function $J$ follows the procedure
outlined in~\cite{news,reduced,bsglrw05}.  As with the hypersurface
equations, we synchronize the fields $\phi$ and $J$, i.e. we communicate
the ghost zone information from each patch to their neighbors, immediately
after advancing radially these two fields with their respective evolution
equations.

\subsection{Boundary treatment for the evolved fields}
\label{subsec:bc}

For the ingoing formulation, we set the field values
$\phi(v,r=r_{out})=0$, and we march inwards until a few points beyond the
black hole horizon ($r=2M$).  Since the past light cones tilts outwards
once inside the horizon, the values computed just inside the horizon
can never affect those points of the grid that lie outside. This scheme
provides then an extremely simple and effective form of excision, as
discussed in~\cite{marsa,wobble,stable} in the context of characteristic
evolution, and in~\cite{bsg} in the context of 3+1 simulations in the
Bondi-Sachs gauge.

For the outgoing formulation on a non--compactified radial grid, we
use simple extrapolative boundary conditions at the outermost point,
i.e.the field $\chi$ at the last point is set equal to the value of
$\chi$ at the point immediately before. This approximation is justified
for sufficiently large $r$ as the field $\phi$ behaves, to leading
order, as $\phi \sim O(1/r)$.  Our treatment of the inner boundary is
motivated by physical considerations that arise naturally in the study
of quasi-normal modes. It can be seen from Eq.~(\ref{eq:potential})
that when the potential $\hat{V}(r)$ goes to zero, as it does in the
limits $r\rightarrow 2M$ and $r\rightarrow \infty$, the solutions
to Eq.~(\ref{eq:qnm}) are traveling waves, $\chi_-=F_L(t+r_*)$ and
$\chi_+=F_R(t-r_*)$. In the linear approximation then, it is consistent
to apply an {\it open boundary} condition to the the scalar field $\phi$
based on the assumption that, at the inner boundary, the field behaves as
a left-travelling wave, $\chi = F_L(t+r_*)$.  It follows also that the
same condition must be applied to the spin-weight 2 metric field $J$. In
the linear approximation, Eq.~(\ref{eq:J}) reduces to Eq.~(\ref{eq:qnm}),
with the potential $\hat{V}(r)$ corresponding to that of a spin--weight
$2$ field, see~\cite{nollert}.  This open boundary condition is equivalent
to stating that the fields $\chi=r\phi$ and $rJ$ propagate towards the
horizon along the incoming characteristics of the two-dimensional wave
operator, Eq.~(\ref{eq:swe}).  In practice, we implement this condition
for the first two points of the radial grid, and use the evolution
equations for $\chi$ and $J$ elsewhere.

In the non-linear case, the horizon can no longer be assumed to be static,
rather it is dynamically distorted and grows as the scalar field accretes
into the black hole. Our boundary condition is applied always to the same
set of points, which are subsequently enveloped by the growing horizon,
thus any inaccuracy we might have introduced in those first two points
can not have any effect on the exterior spacetime.

Our approach suggests the following iterative method to treat the
inner boundary, in a manner which is consistent with the {\it open
boundary} condition: (1) as a first approximation, solve the homogeneous
equation~(\ref{eq:qnm}) for the first two radial points, i.e. assume
the evolved fields propagate along incoming characteristics up to the
retarded time $u+\Delta u$, and (2) with the values predicted for the
fields at time $u+\Delta u$, correct the right-hand side of the full
evolution equations.

\subsection{Tests of second order convergence}

The simulations for the tests reported in the remainder of this section
are conducted in compactified outgoing (retarded) null coordinates.
To verify that the numerical algorithm is globally second-order
convergent, we compute the $L_2$ norm of the relative residuals for
three grid sizes, e.g.
\begin{eqnarray}
{\mathcal Q}_{cm} &=& \int [\chi_c  - \chi_m]^2 dx\,d\Omega,
\nonumber \\
{\mathcal Q}_{mf} &=& \int [\chi_m  - \chi_f]^2 dx\,d\Omega,
\label{eq:Qcmf}
\end{eqnarray}
where the $c$, $m$ and $f$ subscripts denote the field as computed
on coarse, medium and fine grids, respectively.  The field is evolved
from an initial retarded time $u=0$ and the integrals~(\ref{eq:Qcmf})
are calculated at the same final retarded time $u$, using the same
set of spatial grid points, obtained by appropriately subsampling from
the fine and medium grids to the coarse grid. Here we take the angular
(and radial) grids to be in a proportion of $1:3:5$. Grids in these
ratios have a common set of points that align directly, and thus do
not require interpolating cell values from the finest to the coarser
grids. In this case, given the values ${\mathcal Q}_{cm}$ and ${\mathcal
Q}_{mf}$, it can be shown that the order of convergence $O(\Delta^n)$
of the algorithm can be read by solving for $n$ the following equation
\begin{equation}
\left( {\mathcal Q}_{cm}/{\mathcal Q}_{mf}\right)^{1/2}
=
\frac{1 - 1/3^n}{1/3^n - 1/5^n} \ .
\label{eq:order}
\end{equation}
For this test we evolve the initial data
\begin{equation}
\chi(0,r,x^A) = \lambda e^{-(r-r_0)^2/\sigma^2} Y_{\ell m},
\label{initial_data}
\end{equation}
with $\lambda=10^{-4}$, whose the radial profile is characterized by
$r_0=3M$, $\sigma=\frac{1}{2}M$, and whose angular dependence is given by
$\ell=4$ and $m=2$, from $u=0$ up to $u=1M$. We perform three simulations,
on the angular grid sizes $N_\zeta=10,\,30,\, 50$ and the corresponding
radial grid sizes $N_x=501,\, 1501,\, 2501$, for which we take $152, \,
456,\, 760$ time-steps, respectively. From~(\ref{eq:order}) we find that
the measured order of convergence is $n=2.05$, in excellent agreement
with the expected second-order convergence. It should be noted that this
procedure tests the Cauchy convergence of the code, providing a basic
check of the consistency of the discretization. For low amplitudes (in
the perturbative regime), and for a given value of $\ell$, the scalar
field profiles computed with the fully three-dimensional code match,
to within second order, the profiles obtained with a purely radial code
which solves Eq.~(\ref{eq:sph}), as expected.

We want to stress that the boundary conditions, the initial data and the
marching algorithm for the scalar field used in this numerical test are
all the same as those which we have used to calibrate the radial code,
the solutions of which we use here in place of an analytic solution. In
fact, the convergence rate for the radial code is exactly $2.00$ for
the radial grid sizes of $N_x=501, 1501, 2501$, measured at $u=1 M$ with
its respective subsampling, as per Eqs.(\ref{eq:Qcmf})-(\ref{eq:order}).

For sufficiently low values of $(\ell,m)$, the angular grid sizes
$N_\zeta$ used in the convergence test are adequate. For a given angular
grid size, it is also possible to reduce the angular error by increasing
the order of the angular derivatives, for example, to fourth order or higher.
The increased computational expense is offset by the increased
accuracy obtained; in a parallel application there is also the potential
for additional overhead because more ghost cells must be communicated. In
practice we observe that, for the smallest angular grid size considered
($N_\zeta=10$), changing the discretization of the angular derivatives
from second to fourth-order increases the execution time by about 20\%.

In the work reported here, since the radial and time integration are
carried out with a scheme that is second-order convergent, we have
opted to use second order accurate angular derivatives, as with this choice
the nonlinear code exhibits second-order accurate Cauchy convergence.

For the initial data considered here, the radial resolution must
be at least $N_x=501$ to guarantee second-order convergence, as the
radial features are the dominant source of numerical error. We note
that if we repeat the test above using the same angular grid sizes,
$N_\zeta=10,\,30,\, 50$, but using instead radial grids with fewer points,
i.e. $N_x=251, 751, 1250$, the measured convergence rate is lower,
namely $n=1.56$.

For the numerical simulations we present in the remainder of this article
we have chosen grid sizes such that the numerical algorithm is always in
the second-order convergence regime. For more details on the convergence
properties of the radial evolution algorithm, see ~\cite{disip}.

\begin{figure}[!ht]
\includegraphics[width=3.25in,height=3.25in,angle=0]{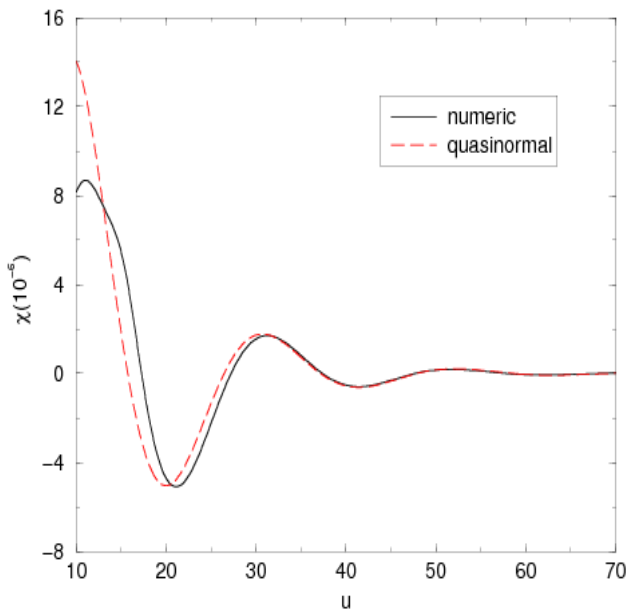}
\caption{The function $\chi(u)$ at $\mathscr{I}$ as a function of Bondi
time, showing the quasi-normal mode regime oscillations for $\ell=1$,
$m=0$. The solid line is the output from LEO for $N_\zeta=11$
and $N_x=1001$, when the initial data and boundary conditions are given
as for the convergence test, the dashed line is the quasi-normal mode
extracted from the data.}
\label{fig:qnm_ell_1}
\end{figure}

\section{Numerical results}
\label{sec:results}

\subsection{Quasi-normal Modes}
\label{subsec:qnm}

\begin{figure}[!ht]
\includegraphics[width=3.25in,height=3.25in,angle=0]{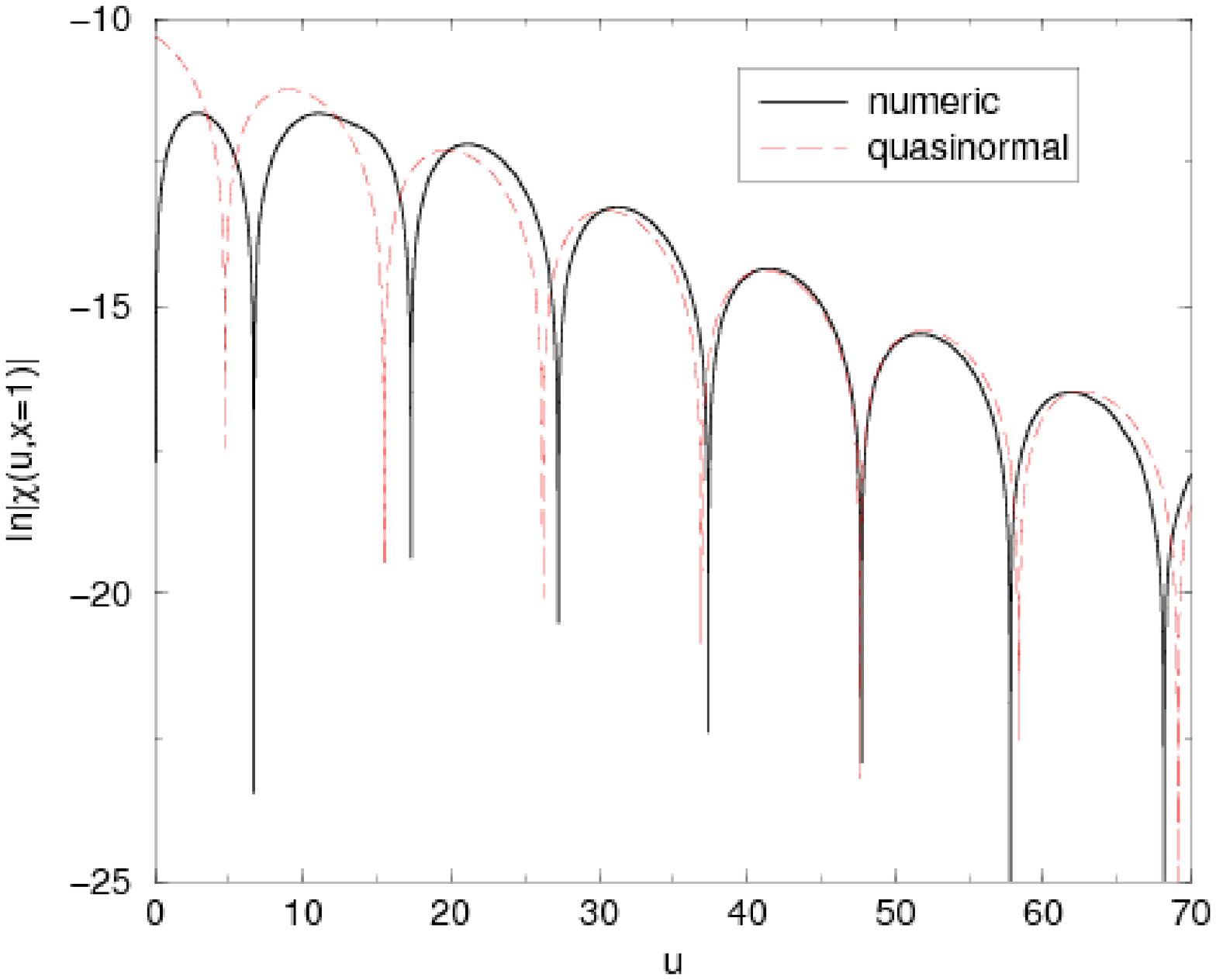}
\caption{Log of the absolute value of the function $\chi(u)$ at $\mathscr{I}$
as a function of Bondi time. Parameters and conditions are the same of
Fig.~\ref{fig:qnm_ell_1}. The solid line is the output from LEO,
dashed)line is the quasi-normal mode extracted the data.}
\label{fig:qnm_ell_1_log}
\end{figure}
\begin{figure}[!ht]
\includegraphics[width=3.25in,height=3.25in,angle=0]{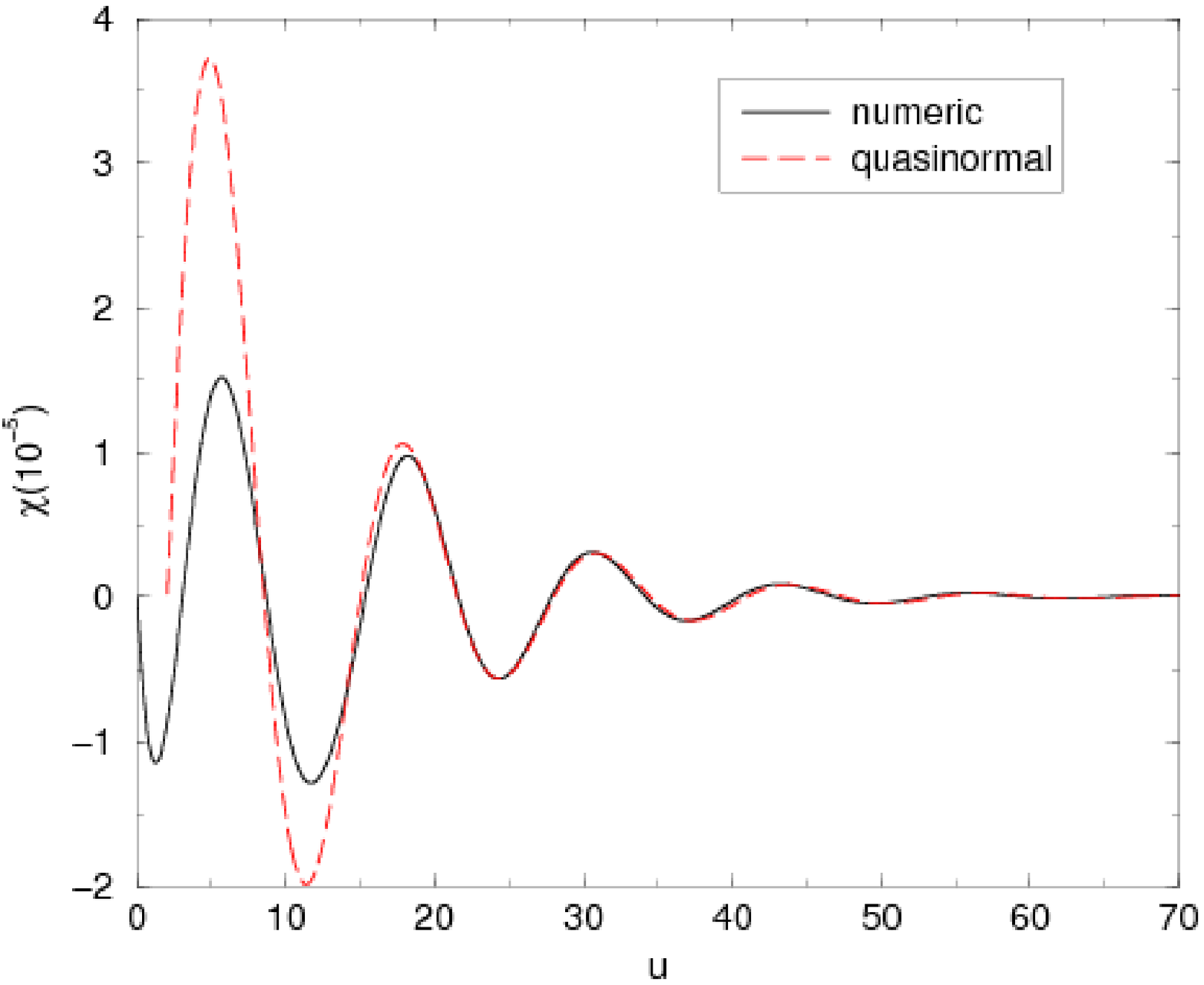}
\caption{The function $\chi$ at $\mathscr{I}$ as a function of Bondi
time, showing the quasi-normal mode regime oscillations for $\ell=2$,
$m=0$. The solid line is the output from LEO for $N_\zeta=11$ and
$N_x=1001$, when the initial data and boundary conditions are given as
for the convergence test, except that $r_{in}=2.13 M$; the dashed
line is the quasi-normal mode extracted from the data.}
\label{fig:qnm_ell_2}
\end{figure}
\begin{figure}[!ht]
\includegraphics[width=3.25in,height=3.25in,angle=0]{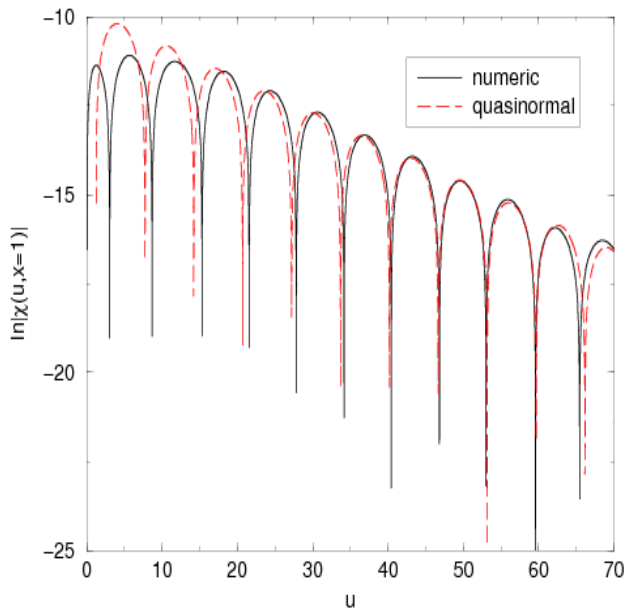}
\caption{Log of absolute value of the function $\chi(u)$ at $\mathscr{I}$
as a function of Bondi time. Parameters and conditions are the same
of Fig.~\ref{fig:qnm_ell_2}. The solid line is the output from
LEO, the dashed line is the quasi-normal mode extracted from the data.}
\label{fig:qnm_ell_2_log}
\end{figure}
The simulations reported in this and subsequent sections are all
carried out in compactified, outgoing (retarded) null coordinates.
Because these coordinates allow us to read off scalar radiation patterns
at null infinity.  (The treatment of the inner boundary is as described
in Sec.~\ref{subsec:bc}.)  The quality of the waveforms extracted depends
in part on the location of the inner boundary and other factors. We
describe here the method used and analyze the sources of error.
For the simulations in this section we use a grid with sizes $N_x=1500$,
$N_\zeta=11$; the initial data corresponds to Eq.~(\ref{initial_data}),
with $\lambda=10^{-4}$, $r_0=3M$, $\sigma=\frac{1}{2}M$, and $M=1$.

To extract the quasi-normal modes we have used the free software
package Harminv~\cite{harminv}, which employs a low-storage filter
diagonalization method (FDM) for finding the quasi-normal modes in
a given frequency interval. This software package is based on the FDM
algorithm described in \cite{mt97,mt98e}. The advantage of using Harminv
is that FDM methods provide better accuracy than what can be obtained
with a fast Fourier transform (FFT)~\cite{mode}, and are more robust
than least-squares fit~\cite{dorband}. We find it surprising that this
approach, to our knowledge, has not been used in the context of reading
quasi-normal modes in gravitational simulations.

In performing a fit with Harminv to the scalar field waveforms,
we find sometimes necessary to factor out, at least approximately,
the exponential decay of the signal. This happens when the magnitude
of the imaginary part of the frequency (the decay rate) is comparable
to the real (oscillatory) part, where the FDM method fails to find a
fitting frequency. In those cases, we pre-multiply the signal by an
exponentially increasing function $f=\exp( |\omega_f| t)$, perform the
fit with Harminv, and adjust the frequency obtained accordingly. When
an analytic value for the frequency is available, we take its imaginary
part as the value for $\omega_f$. In general, when the imaginary part
of the frequency is not known, it suffices to use a rough estimate of
the decay rate, which can obtained graphically.
We also need to decide what range of values of $u$ to use to extract
this information. We do this by plotting the signal $\chi(u)$ and
noting when the waveform is clearly periodic with an exponentially
decaying envelope. For example, in Fig.~\ref{fig:qnm_ell_1_log}, one
can clearly see that this regime starts at about $u=20 M$. We take the
end of the fitting interval when the signal no longer appears to be a
damped sinusoidal. For initial data of the form~(\ref{initial_data}),
with $\ell=1$, we use Harminv to extract the frequency, using as
the fitting interval $u=\left[20,70\right]$. The measured frequency
is $\omega= 0.3076\, (5\%) -0.1064\,i \,(9\%)$. Here the values
in parenthesis indicate the percentage deviation from the value
calculated in~\cite{konoplya} via the WKB method to sixth order. A
comparison of the signal computed and the quasi-normal mode fitted is shown
in Figs.~\ref{fig:qnm_ell_1}-\ref{fig:qnm_ell_1_log}. 
The figures show the profiles computed with the three-dimensional code
(solid line). These profiles are indistinguishable, at the resolution
of the graph, from the profiles obtained by solving numerically the
perturbative equation (\ref{eq:sph}) for the same initial data, thus we
have opted not to show the perturbative solution as is customary. For
comparison, we have shown instead, in the same graph, the quasi-normal mode
$\chi=\exp{\omega u}$ (dashed line) extracted, i.e. the fit provided by Harminv.
There is some disagreement initially between the numerical solution and
the fit, as the numerical solution settles into the dominant quasi-normal
mode, a process which takes from one to one and half cycles of the quasi-normal mode.

For the same initial data, but with $\ell=2$, we read a
frequency $\omega=0.4971\, (3\%) -0.0992 \,i\, (2\%)$,
in the range $u=\left[40,70\right]$, with the comparison
between the computed signal and~\cite{konoplya} shown in
Figs.~\ref{fig:qnm_ell_2}-\ref{fig:qnm_ell_2_log}. We have observed that
the relative percent error for the decay rate is larger for $\ell=1$
because it depends strongly on the value selected for the location
of the boundary, $r_{in}$. Numerical experiments with the radial code
confirm this and suggest that, by carefully tuning the location of the
inner boundary, better accuracy can be achieved for any one value of
$\ell$. We have done this only partially in computing the frequency
for the case $\ell=2$. We want to emphasize that the dependence of the
frequency on the boundary is not a numerical artifact of the code, but a
consequence of the choice of outgoing null coordinates. This is confirmed
by numerical experiments with the radial code in ingoing coordinates,
in which case we find that the frequency can be read off with an error
of less than $0.1\%$ for the same initial data and grid sizes.

\subsection{Energy Conservation}
\label{subsec:energycons}

\begin{figure}[!ht]
\includegraphics[width=3.25in,height=3.25in,angle=0]{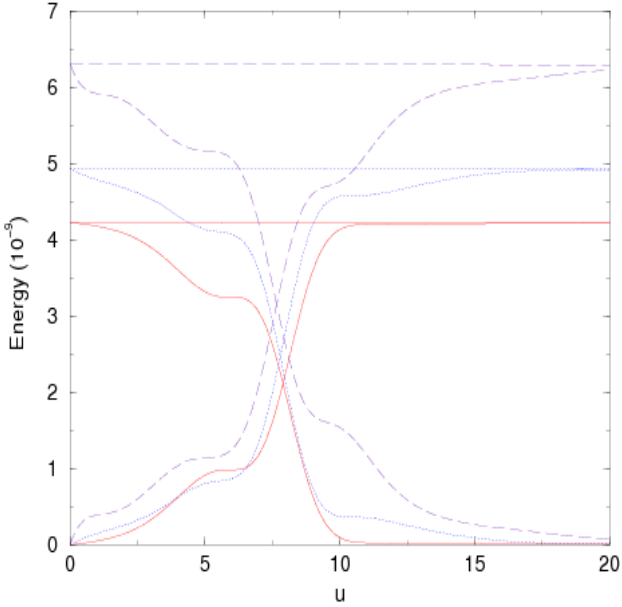}
\caption{Energy conservation as a function of Bondi time for $\ell=0$
(solid line); $\ell=1$ (dotted line); $\ell=2$ (long
dashed line). This calculation was done using the same grid parameters as
for Fig.~ \ref{fig:qnm_ell_2} except for $r_{in}=2.3$. For each specific
$\ell$ (line type; color) the descending curve corresponds to energy
given by Eq.~(\ref{eq:econs}). The ascending curve corresponds to the
algebraic sum of $E_{in} = -\int P_{in} du$ and $E_{out}=\int P_{out}
du$. Thus, in accordance with Eq.~(\ref{eq:gecons}), the horizontal
curve represents the global conservation of energy.}
\label{fig:energy}
\end{figure}
\begin{figure}[!ht]
\includegraphics[width=3.25in,height=3.25in,angle=0]{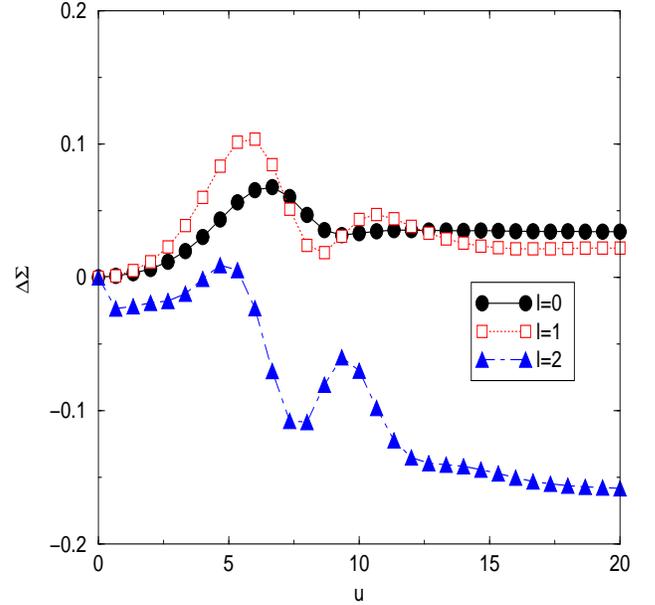}
\caption{Percentage variation in $\Sigma(u)$ with respect to $\Sigma(0)$
as a function of Bondi time for $\ell=0$ (circles), $\ell=1$ (squares),
and $\ell=2$ (triangles). The graph shows that energy is conserved to
within less than $0.2\%$ of the energy content of the initial surface.}
\label{fig:energy_percentage}
\end{figure}
\begin{figure}[!ht]
\includegraphics[width=3.25in,height=3.25in,angle=0]{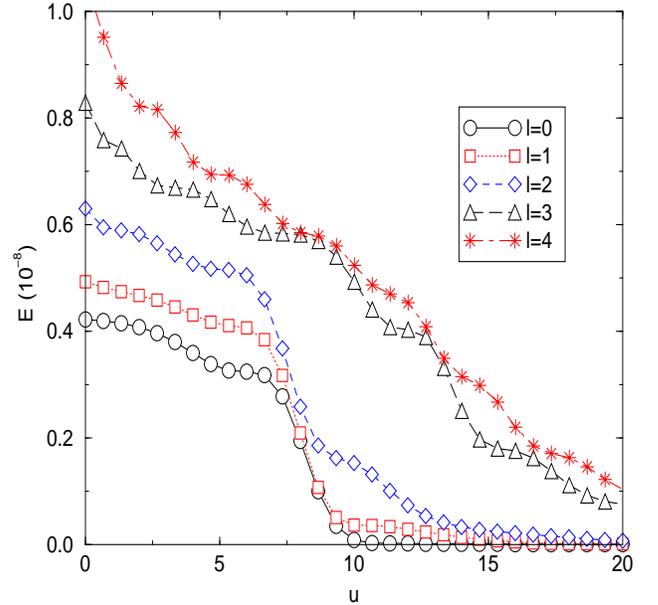}
\caption{Energy content $E(u)$ as a function of Bondi time for: $\ell=0$
(circles), $\ell=1$ (squares), $\ell=2$ (diamonds), $\ell=3$
(triangles), and $\ell=4$ (stars). This calculation was
done using the grid parameters $N_\zeta=11$ and $N_x=1001$. The initial
data and boundary conditions are the same as in the convergence test.}
\label{fig:energy_III}
\end{figure}
\begin{figure}[!ht]
\includegraphics[width=3.25in,height=3.25in,angle=0]{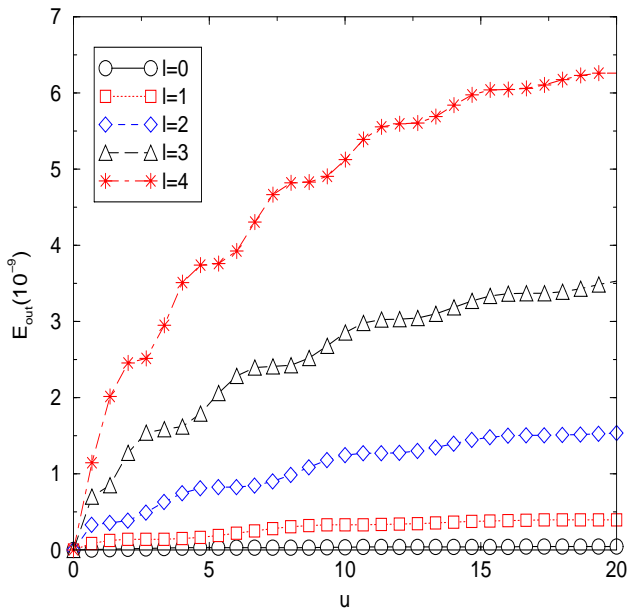}
\caption{Energy flow to infinity $E_{out}=\int P_{out}(u) du$ as a
function of Bondi time for: $\ell=0$ (circles), $\ell=1$ (squares), $\ell=2$
(diamonds), $\ell=3$ (triangles), and $\ell=4$ (stars). This calculation
was done using the same conditions of Fig. \ref{fig:energy_III}.}
\label{fig:energy_out}
\end{figure} 
\begin{figure}[!ht]
\includegraphics[width=3.25in,height=3.25in,angle=0]{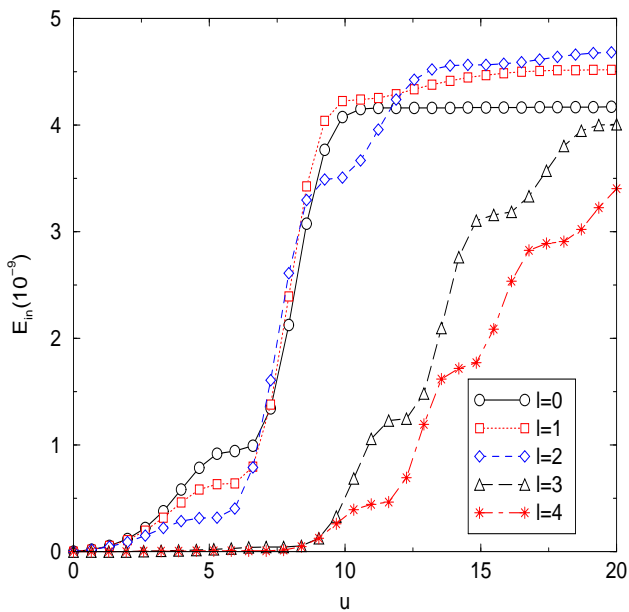}
\caption{Energy flow into the black hole, $E_{in}=-\int P_{in}(u)
du$, as a function of Bondi time for: $\ell=0$ (circles), $\ell=1$
(squares), $\ell=2$ (diamonds), $\ell=3$ (triangles), and $\ell=4$
(stars). Both curves for $\ell=3$ and $\ell=4$ saturate eventually
without crossing for $u>30$. This calculation was done using the same
conditions of Fig. \ref{fig:energy_III}. }
\label{fig:energy_in}
\end{figure} 
For initial data of the form~(\ref{initial_data}) with $\ell=0,1,2$,
Fig.~\ref{fig:energy} shows that energy is conserved in the linear
regime. It is immediately clear from the graph that the energy contained
on the initial slice is larger the larger the value of $\ell$. In all
cases energy is clearly conserved, however, we have seen also that if
the resolution is not sufficient for a given $\ell$, this fact shows
up clearly in the graph of energy conservation. Thus, we can use energy
conservation, as well as the results from running the same initial data
on the radial code, to debug and calibrate the nonlinear code, as well as
to estimate the evolution time needed and its computational requirements.
From Fig.~\ref{fig:energy} alone the reader might be left to guess as to
the extent of the deviation of the total energy from a straight line,
since that deviation is clearly so small that it does not show up in
the plot for any of three simulations reported in Fig.~\ref{fig:energy}.
Fig.~\ref{fig:energy_percentage} shows the variation in the energy balance
$\Delta \Sigma(u)$, defined as the percentage variation in $\Sigma(u)$
relative to the initial value, $\Sigma(u_0)$, i.e. 
\begin{equation}
\Delta \Sigma = (\Sigma(u)/ \Sigma(0) - 1) \times 100 ,
\end{equation}
It can be seen from Fig.~\ref{fig:energy_percentage} that the relative
change $\Delta\Sigma(u)$ stays below $0.2\%$ during the simulation. We
will revisit energy conservation in the context of large resolution
simulations in Sec.~\ref{subsec:largeres}.

Fig.~\ref{fig:energy_III} shows the energy content $E(u)$ as a
function of Bondi time $u$ for a sequence of simulations with initial
data~(\ref{initial_data}) with varying values of $\ell$. For lower values
of $\ell$ $(\ell=0,1,2)$, the energy content $E(u)$ decays slowly at
first, then drops rather sharply, and afterwards it decays again slowly,
at a much lower rate. For higher values of $\ell$, $(\ell=3,4)$, the
energy decays approximately monotonically from the beginning of the
simulation.

We also observe, see Fig.~\ref{fig:energy_out}, that in general,
increasing $\ell$ corresponds to an increase of the energy radiated
at $\mathscr{I}$. The oscillations observed in the profiles are
higher the higher the value of $\ell$, as would be expected. The most
interesting observation in the analysis of energy balance arises from
Fig.~\ref{fig:energy_in}, and is the following:  for values of $\ell$
from 0 to 2, the total energy flux towards the black hole (as measured
by $E_{in}(u)$ as $u\to \infty$) increases with the value of $\ell$;
however, for values of $\ell \ge 2$, the total flux of energy towards
the black hole diminishes with increasing values of $\ell$. We have
confirmed that this is the case with the radial code, so this is not a
non-linear effect. It is also clear that the sudden change of energy for
$\ell \le 2$ is due to the energy carried away by the scalar field as
it falls into the black hole. At about $\ell=2$, the radiation into the
black hole saturates, and for higher values of $\ell$, i.e. for $\ell >
2$, the centrifugal potential barrier prevents much of the field from
falling into the black hole. Thus, for the same amplitude, configurations
with higher angular momentum (larger $\ell$ values) carry more energy,
most of which will be radiated away and less of which will fall into
the black hole, so in that sense these configurations are proportionally
more efficient at carrying energy out to $\mathscr{I}$.

\subsection{Large resolution simulations}
\label{subsec:largeres}

\begin{figure}[!ht] 
\includegraphics[width=3.25in,angle=0]{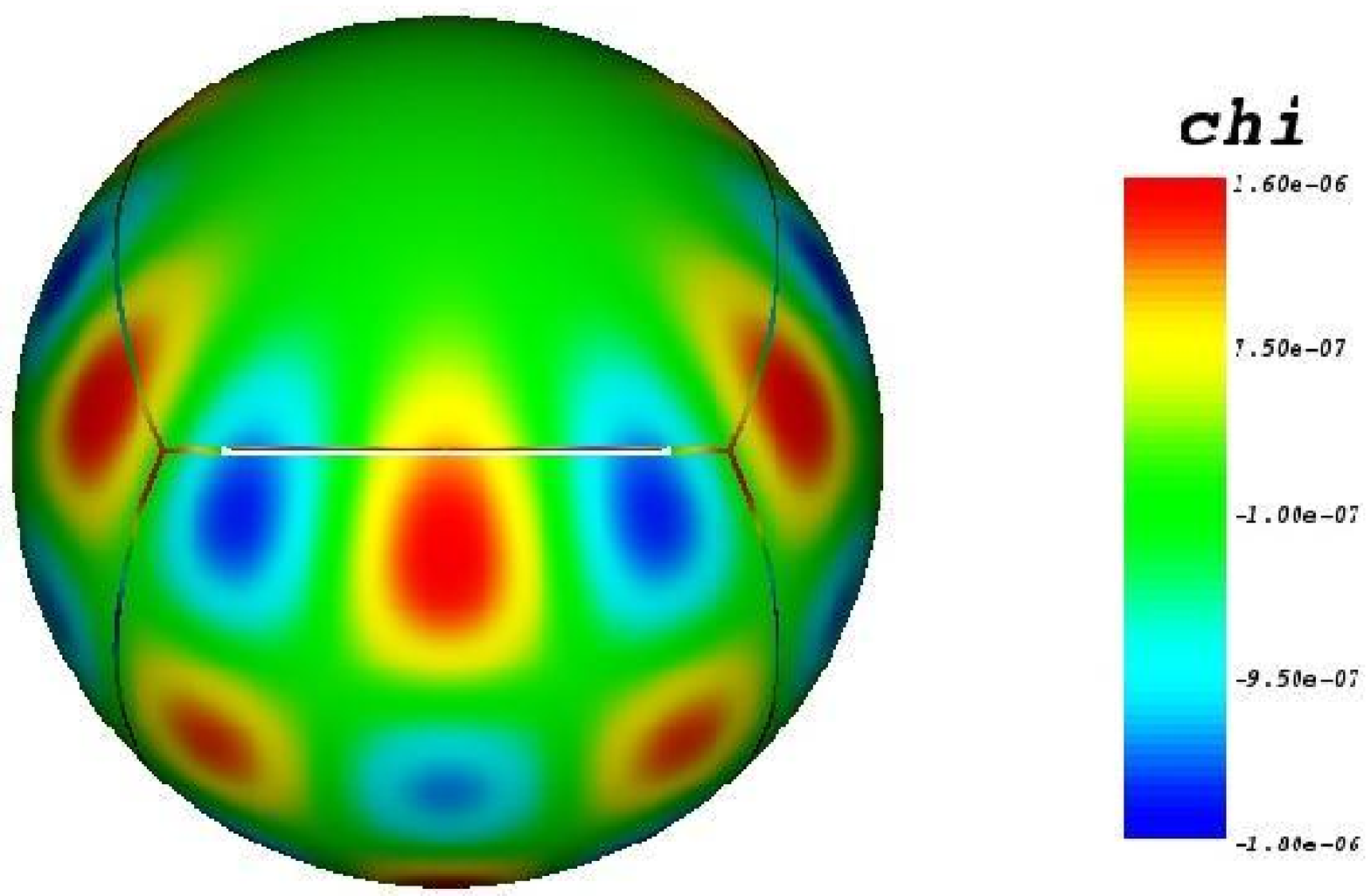}
\includegraphics[width=3.25in,angle=0]{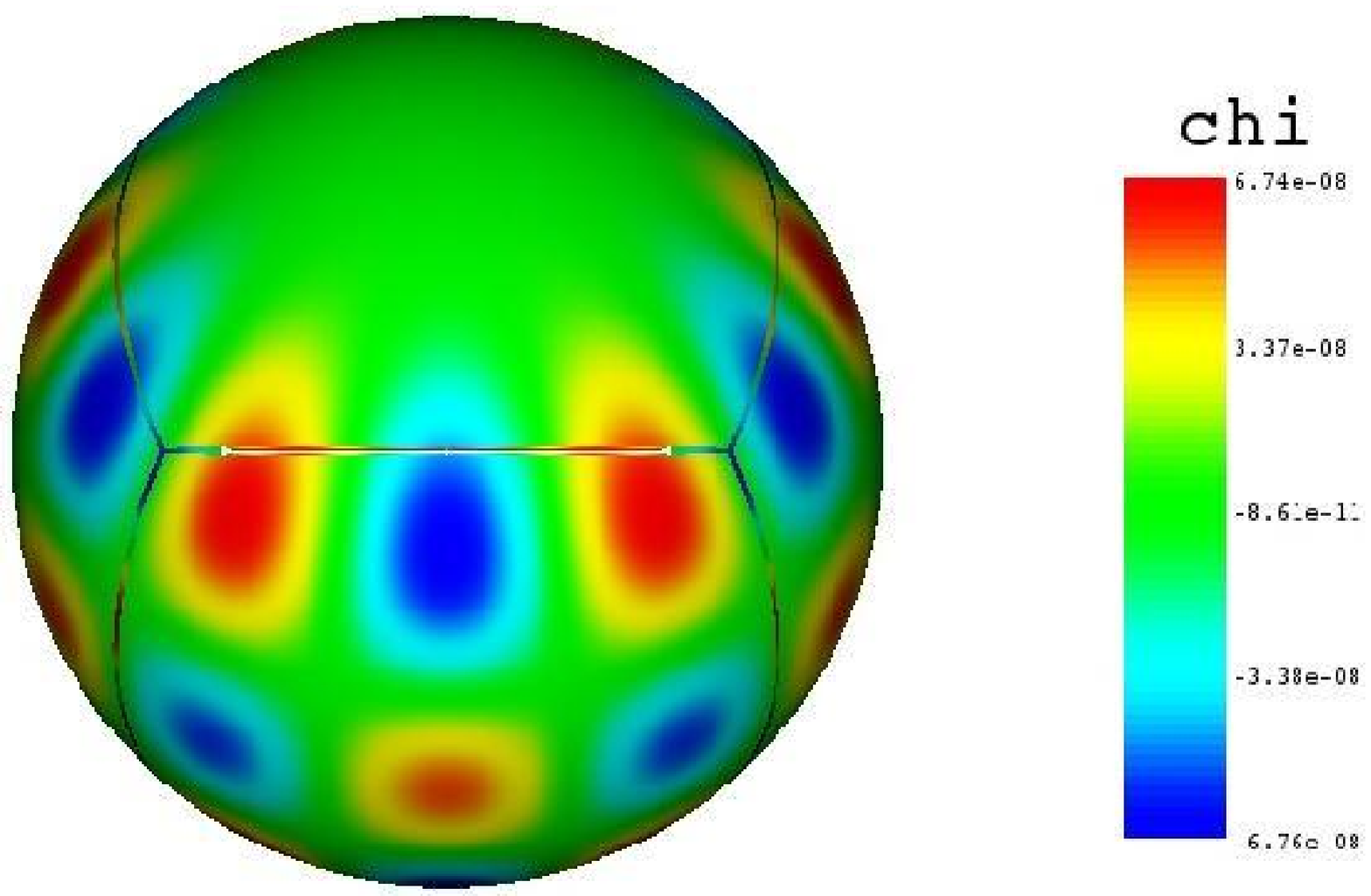} 
\caption{Surface plots of $\chi$ at  $\mathscr{I}$ for $u=2.5M$
(top) and $u=30M$ (bottom).  The parameters of the initial data are
$\lambda=10^{-4}$, $r_0=3M$, $\sigma=0.5M$, $\ell=8$, $m=6$. The grid
size is $N_\zeta=93$, $N_x=1501$.}
\label{fig:phi@scri} 
\end{figure}
\begin{figure}[!ht]
\includegraphics[width=3.25in,angle=0]{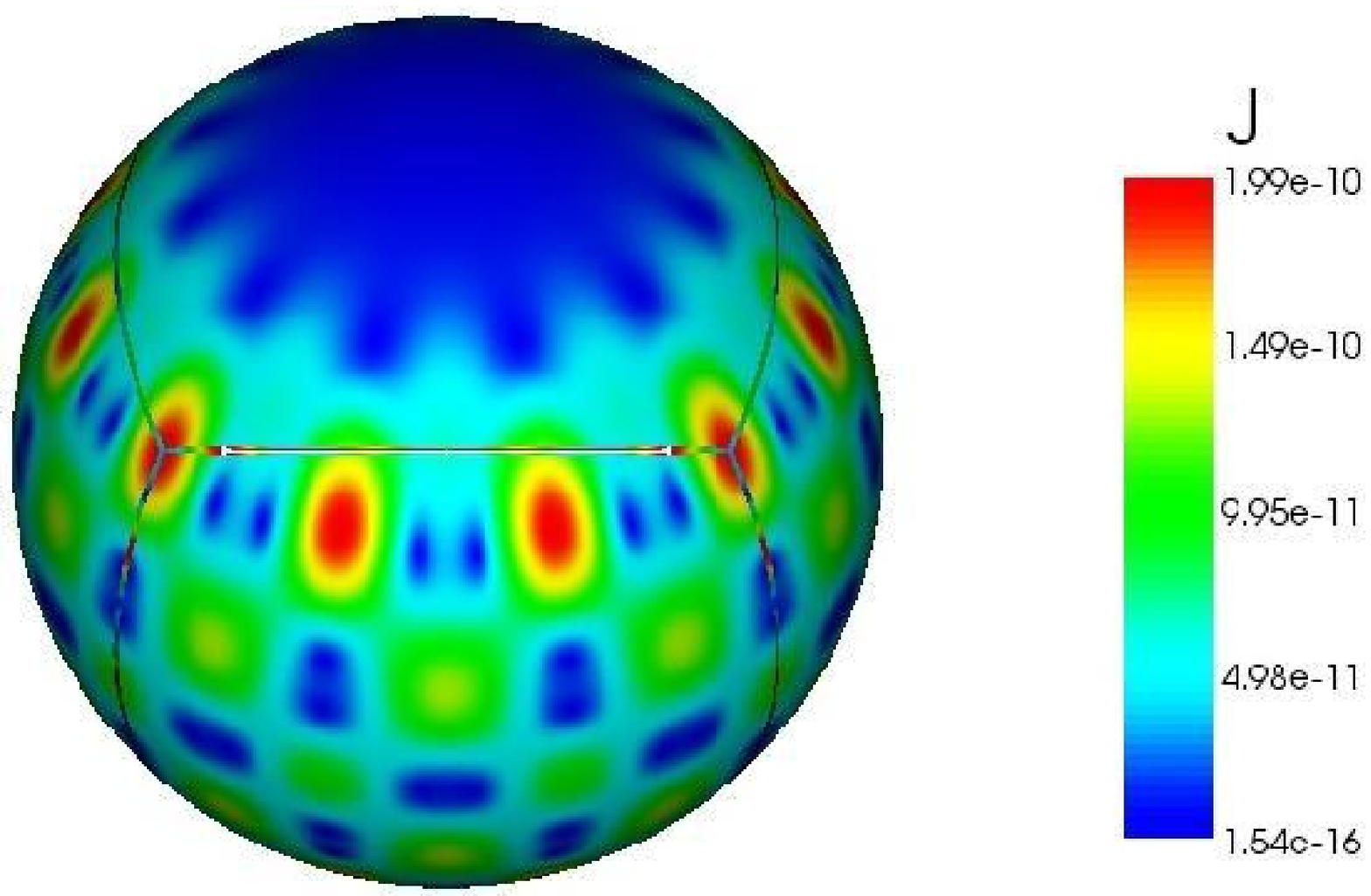}
\includegraphics[width=3.25in,angle=0]{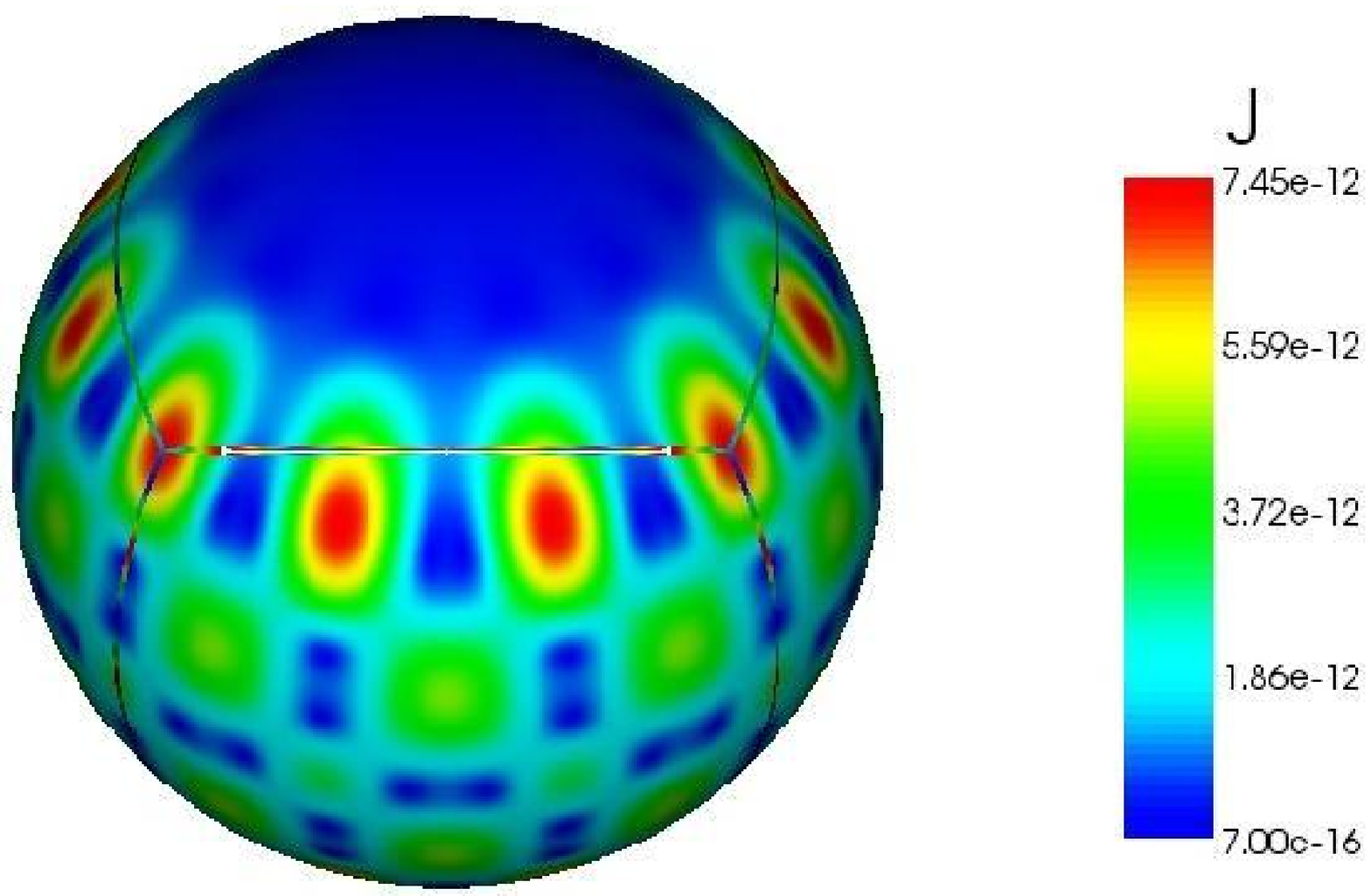}
\caption{Surface plots of $J\bar J$ at $\mathscr{I}$ for $u=2.5M$ (top) and
$u=30M$ (bottom). The parameters of the initial data and grid size are
the same as in Fig.\ref{fig:phi@scri}.}
\label{fig:J@scri}
\end{figure}
\begin{figure}[!ht]
\includegraphics[width=3.25in,angle=0]{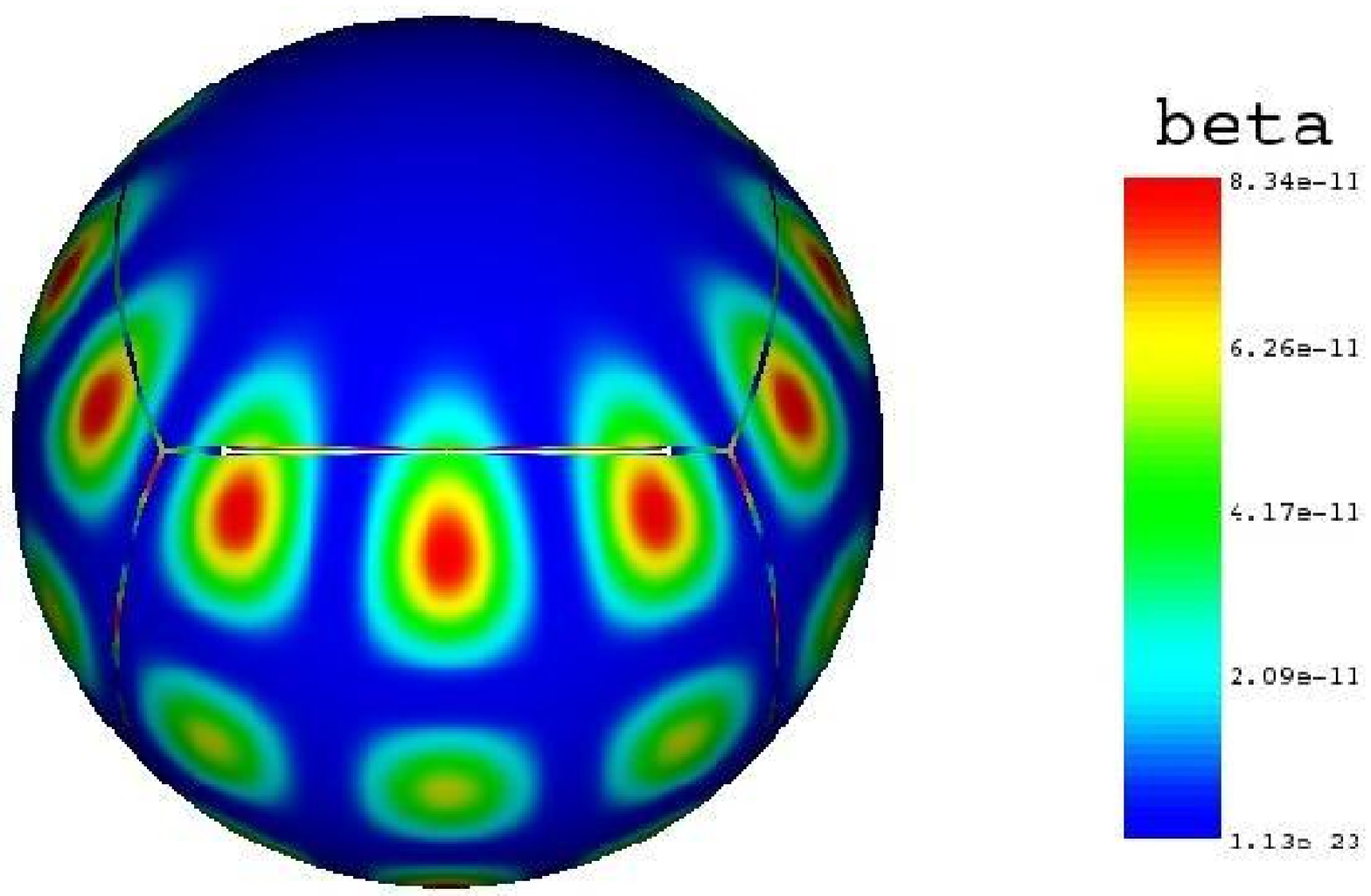}
\includegraphics[width=3.25in,angle=0]{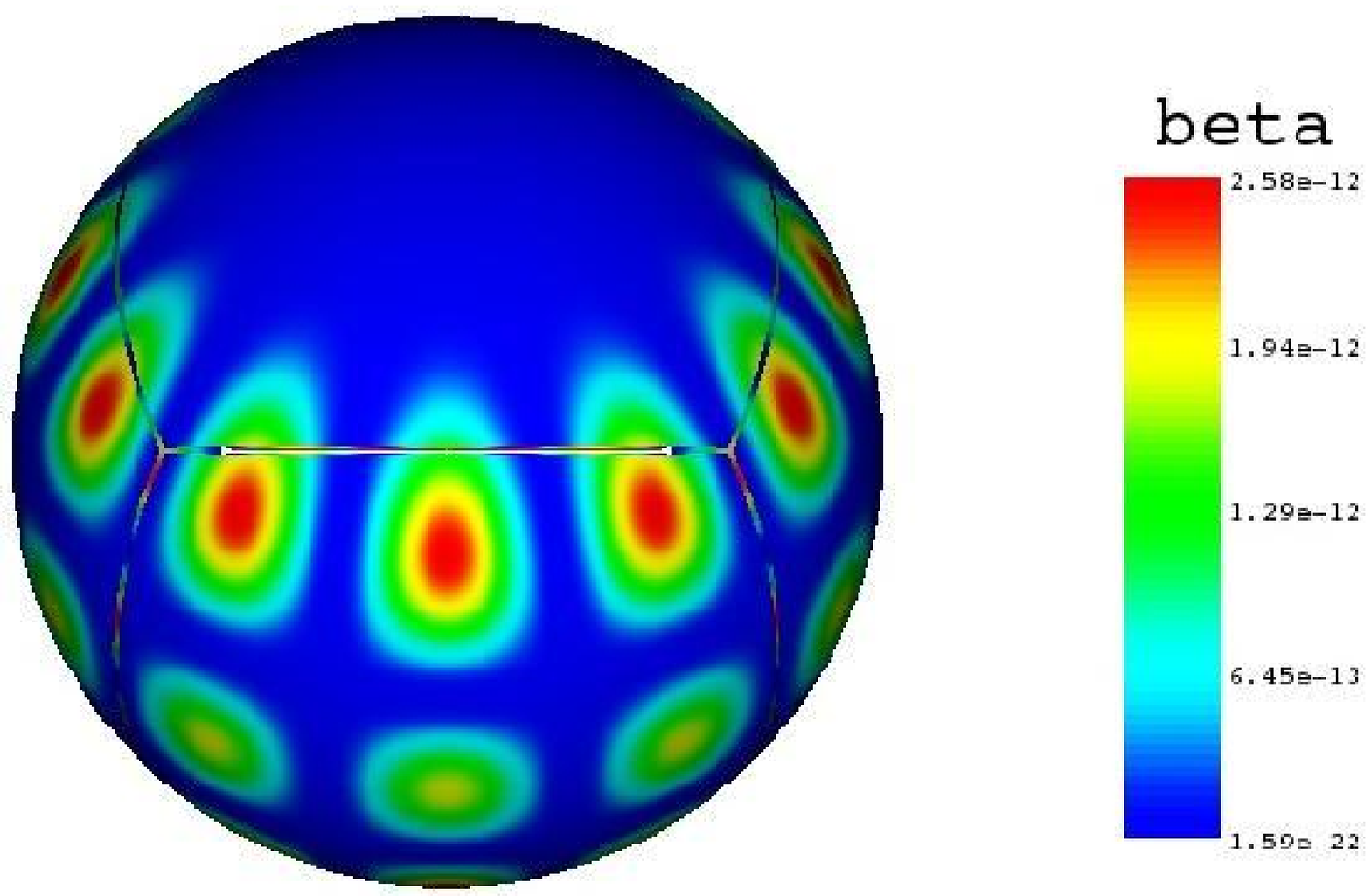}
\caption{Surface plots of $\beta$ at $\mathscr{I}$ for $u=2.5M$ (top)
and $u=30M$ (bottom). The parameters of the initial data and grid size
are the same as in Fig.\ref{fig:phi@scri}.}
\label{fig:beta@scri}
\end{figure}
\begin{figure}[!ht]
\includegraphics[width=3.25in,angle=0]{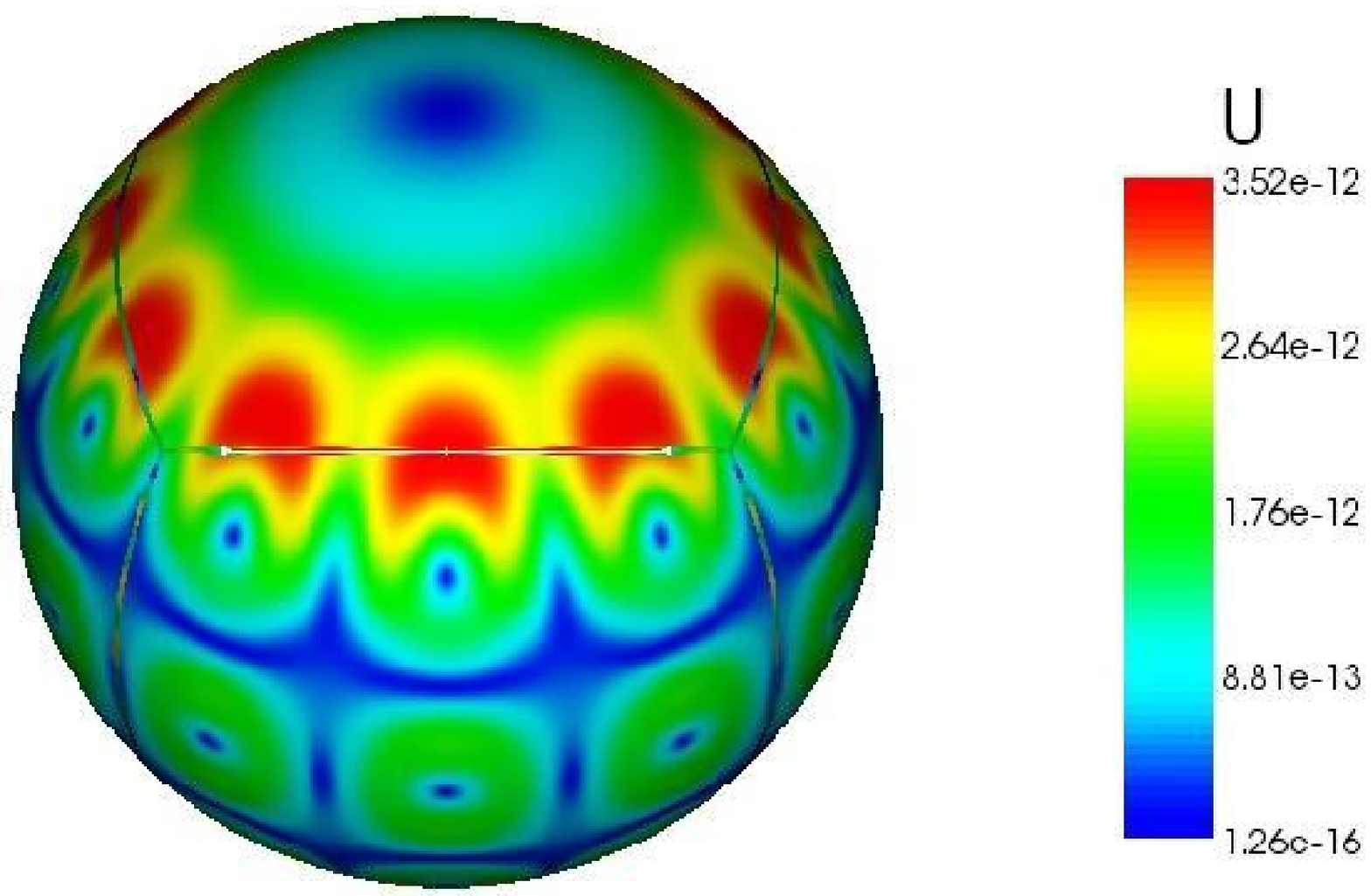}
\includegraphics[width=3.25in,angle=0]{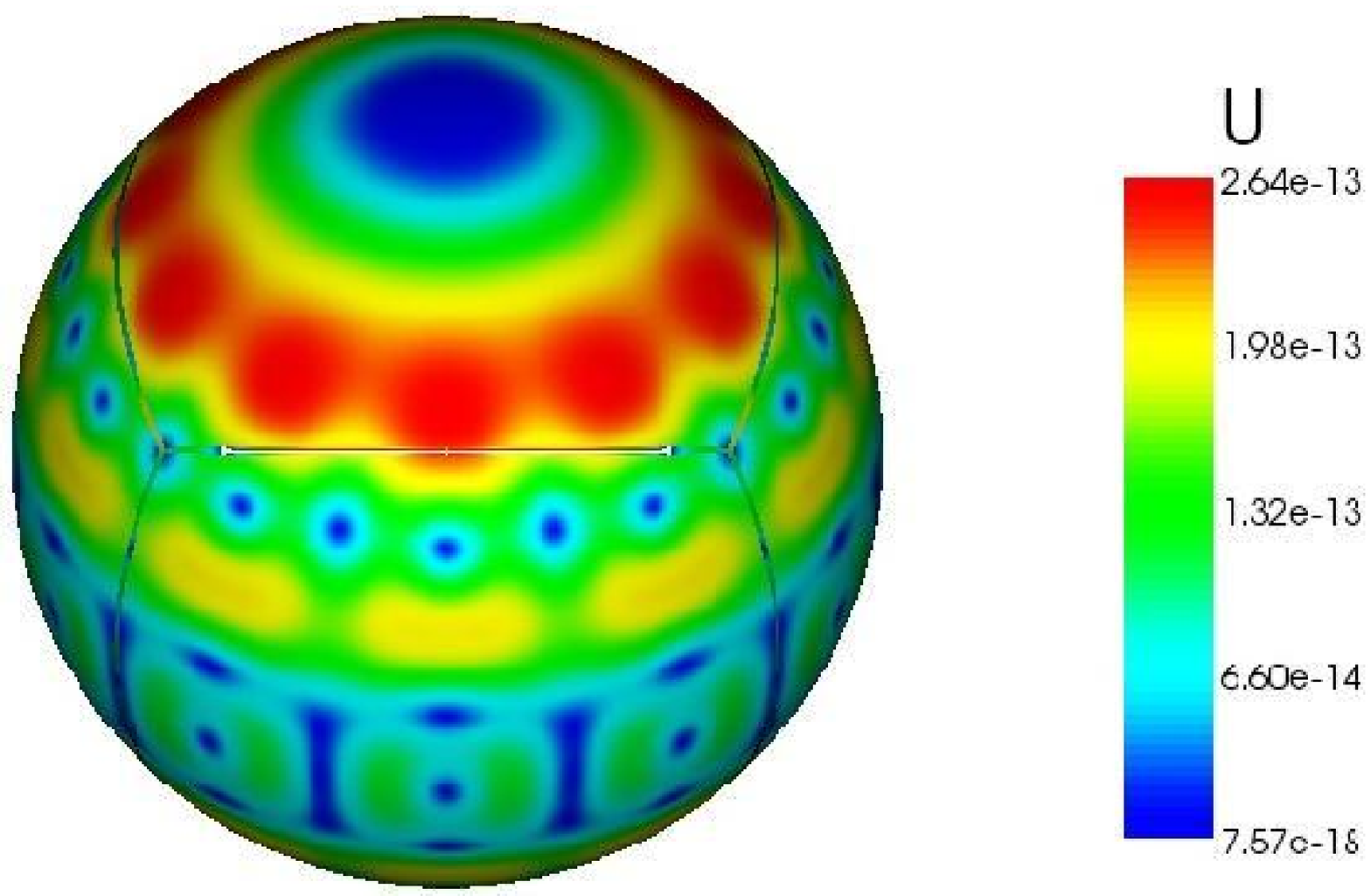}
\caption{Surface plots of $U\bar U$ at $\mathscr{I}$ for $u=2.5M$ (top) and
$u=30M$ (bottom). The parameters of the initial data and grid size are
the same as in Fig.\ref{fig:phi@scri}.}
\label{fig:U@scri}
\end{figure}
\begin{figure}[!ht]
\includegraphics[width=3.25in,angle=0]{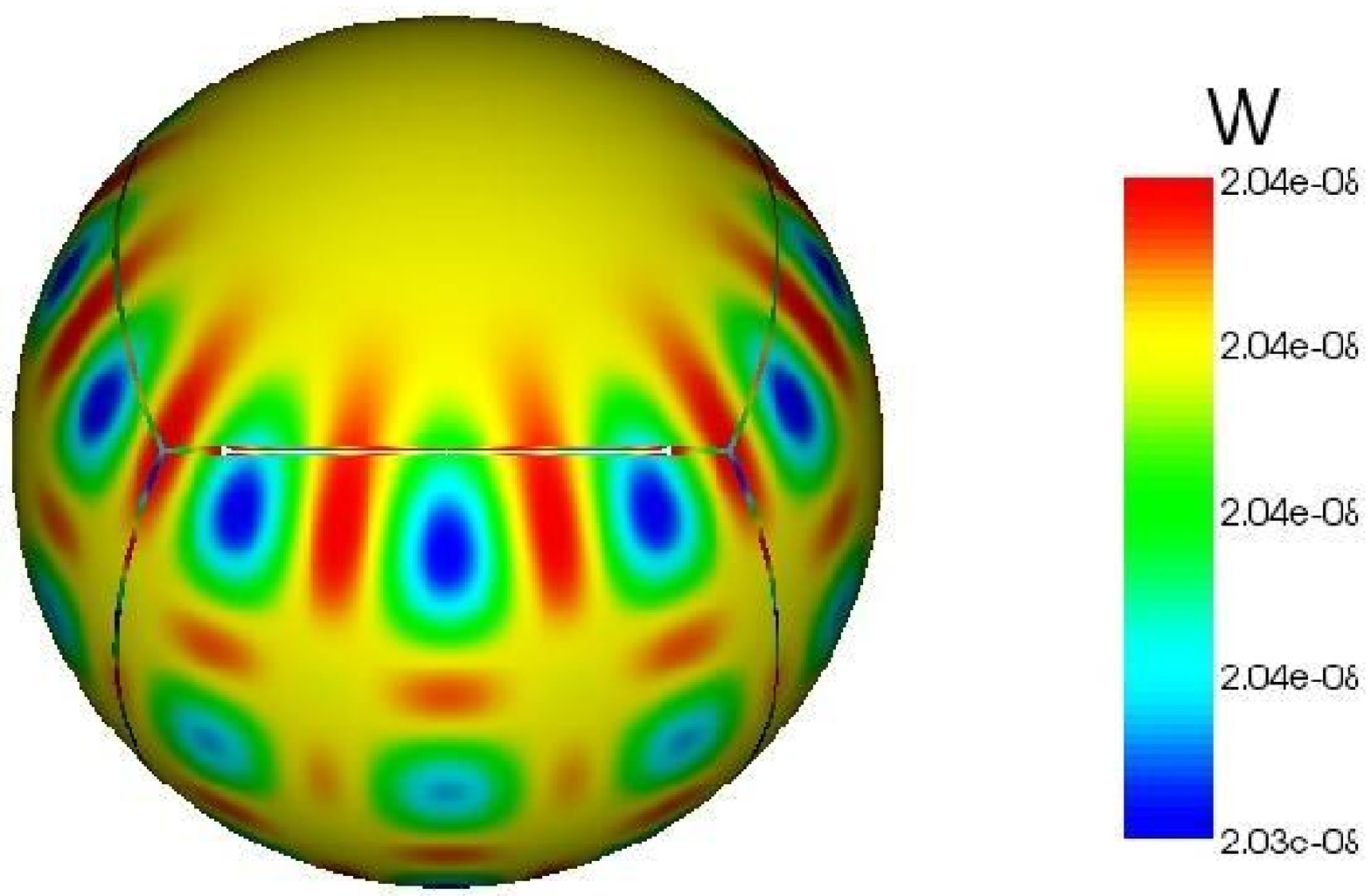}
\includegraphics[width=3.25in,angle=0]{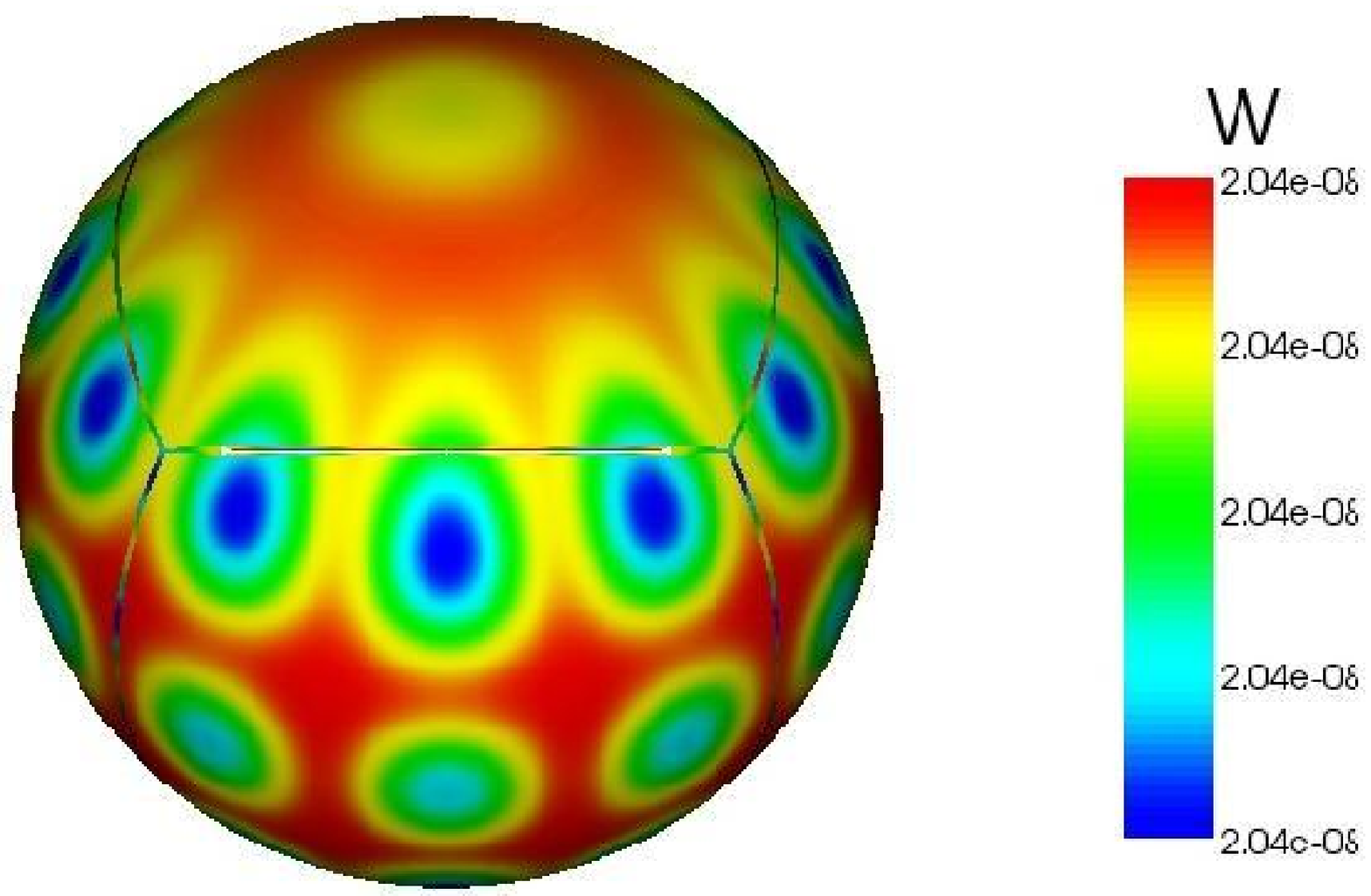}
\caption{Surface plots of $W$ at $\mathscr{I}$ for $u=2.5M$ (top) and
$u=30M$ (bottom). The parameters of the initial data and grid size are
the same as in Fig.\ref{fig:phi@scri}.}
\label{fig:W@scri}
\end{figure}
\begin{figure}[!ht]
\includegraphics[width=3.25in,angle=0]{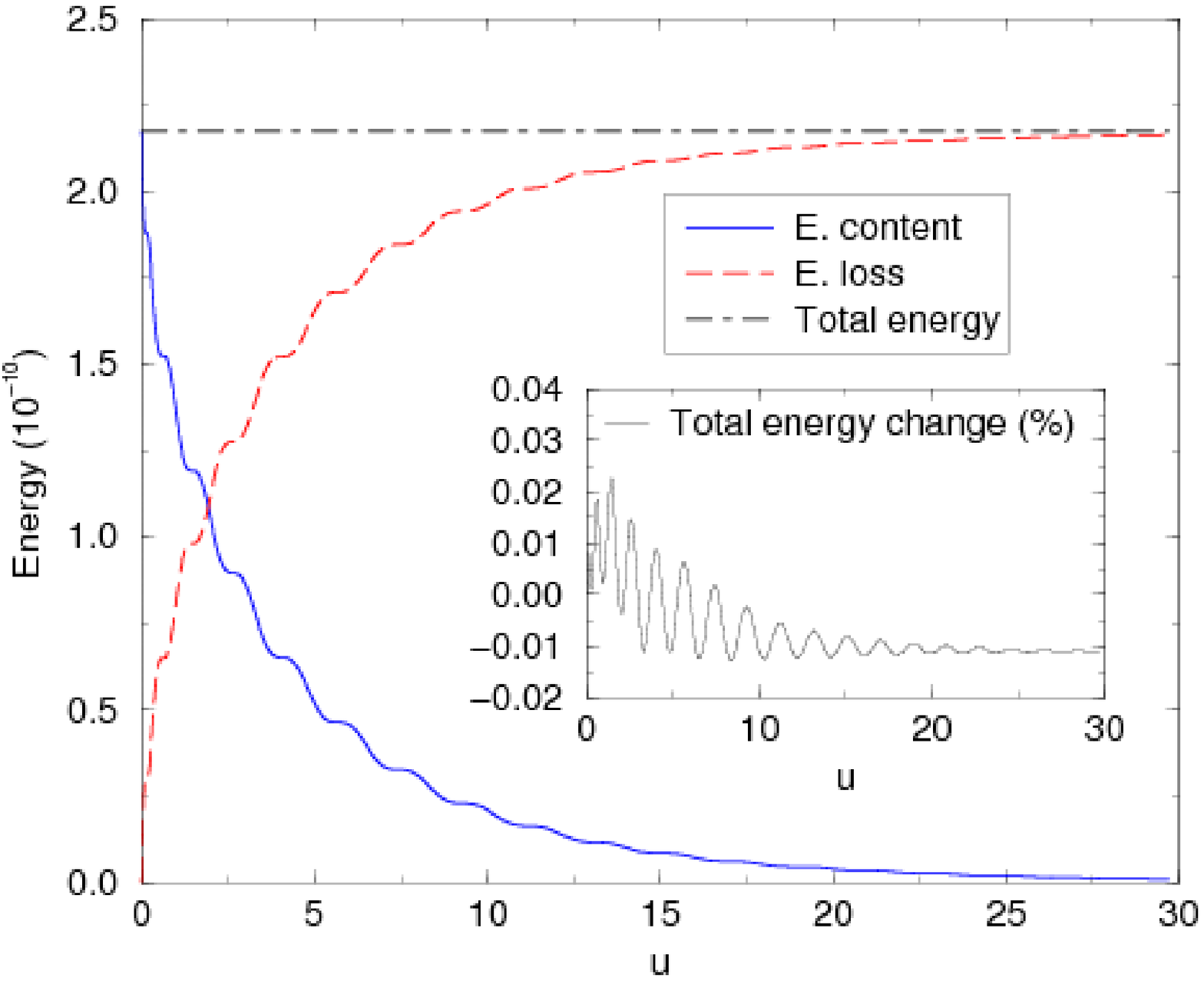}
\caption{Energy conservation for the simulation which generated the
results shown in Figs.~\ref{fig:phi@scri}--\ref{fig:W@scri}. The solid
line corresponds to the energy content $E(u)$ at successive times,
the dashed line to the sum of the energy radiated through the inner
($E_{in}(u)$) and outer ($E_{out}(u)$) boundaries, and the dot-dashed
line to the sum, $\Sigma(u)=E(u)+E_{in}(u)+E_{out}(u)$, respectively.
The insert graph shows the percentage variation in $\Sigma(u)$, relative
to its final value at $u=30$.}
\label{fig:energy_IV}
\end{figure}

In order to get a first glimpse of the type of simulations that
our framework enables us to perform, and to perform a final
calibration check of the nonlinear code, we select initial data
given by Eq.~(\ref{initial_data}), with $\lambda=10^{-4}$, $r_0=3M$,
$\sigma=\frac{1}{2}M$, $\ell=8$, $m=6$, and evolve this configuration until
$u=30 M$.  This simulation is performed in compactified outgoing
coordinates, with the treatment of  the inner boundary as described
in Sec.~\ref{subsec:bc}. The plots shown are of quantities computed at
null infinity,  $\mathscr{I}$.  The angular grid has size $N_\zeta=93$,
that is, there are 372 points on a great circle on the sphere, while the
radial grid has $N_x=1501$ points. This simulation requires 27 hours on
54 processors, for a total of 1458 processor--hours, or the equivalent
of two months of a single-processor run. It is not by far the largest
simulation we could run with our framework: we have performed scaling
studies that indicate the code scales linearly well into the 4000+
processor range, but it suffices as a demonstration of the resolution
that can be achieved and the typical turn--around times. We assign no
particular significance to the initial data selected, other than the
fact that its angular complexity provides an excellent test of the code.
On any large simulation, data analysis and visualization is always a
challenge. In LEO, visualization is performed by having each processor
write its own data set at run time, the individual data files are then
post-processed, and graphs of the desired quantities generated with
Paraview~\cite{paraview}. Paraview allows us to easily generate graphs
of slices at constant coordinate lines and volumetric renderings of
various fields. Of particular interest to us is the behavior of the
various metric quantities at null infinity.

Figs.~\ref{fig:phi@scri}-\ref{fig:W@scri} display the code variables
$\phi$, $J$, $\beta$, $U$ and $W$ as functions on the sphere at null
infinity, at $u=2.5M$ and at $u=30M$. In each case, the graphs show
the corresponding field on the six cubed--sphere caps (the gap between
the caps is the actual size of the spacing between grid cells). The
north pole is at the top of the figure, rotated 45 degrees towards
the observer. For those fields that are complex (and which have
spin different from zero), i.e. $J$, $U$, we display for ease of
visualization the combinations $J\bar J$ and $U\bar U$, which are real
and have spin zero. Clearly visible in Fig.~\ref{fig:phi@scri} is the
$m=6$ azimuthal dependence of the field, marked by the presence of six
maxima and minima. It is also apparent that $\chi$ oscillates in time,
as the maxima and minima alternate between the top and bottom figures.
The angular dependence is preserved by the evolution, as expected,
as the only change between the two figures is in the overall amplitude
(by a factor of $\approx 25$ in between the two times shown). The graphs
of $J\bar J$, Fig.~\ref{fig:J@scri}, are clearly different in their
angular dependence, showing the presence of various harmonics at earlier
time. This can be understood since in our initial data $J(r,x^A)=0$,
thus $J$ develops from $\chi$, i.e. $J \sim (\eth \phi)^2$, and it
is not until later times that a definite profile for $J$ has formed.
The graph of $\beta$, Fig.~\ref{fig:beta@scri}, shows precisely the
angular dependence resulting from the contribution $\beta_{,r} \sim
(\phi_{,r})^2$ to the source term in Eq.~(\ref{eq:beta}), and remains
constant throughout the simulation, up to an overall amplitude. The graphs
of $U$ and $W$, Figs.~\ref{fig:U@scri}-\ref{fig:W@scri}, show higher order
angular dependence arising from the angular derivatives of $U$, which in
turn are essentially driven by the source term in the Eq.~(\ref{eq:Q}),
i.e. by $U \sim Q \sim \phi\eth\phi$. 

Fig.~\ref{fig:energy_IV} shows again that energy is conserved during the
entire simulation. The variation in the energy balance $\Sigma(u)$ is well
below $1\%$, thus $\Sigma(u)$ is indistinguishable from a straight line at
the resolution of the graph, as noted in Sec.~\ref{subsec:energycons}. To
more fully appreciate to what extent energy is conserved, the graph insert
in Fig.~\ref{fig:energy_IV} shows the percentage variation of $\Sigma(u)$,
normalized to its value at $u=0$. From the graph insert we see that
the total energy varies by at most $0.025\%$ during the simulation. As
we stated earlier, energy conservation is a requisite for accuracy in
the waveforms; note that energy is conserved during this simulation to
within tighter limits than in the simulations of Sec.~\ref{subsec:qnm},
which is due primarily to the increased angular resolution. For this
simulation we took $N_\zeta=93$ in order to accurately resolve the
higher order harmonic angular dependence, where in Sec.~\ref{subsec:qnm}
we set $N_\zeta=11$; this amounts to (approximately) an 8-fold increase
in resolution, and for the same radial resolution, a 64-fold increase
on the computing resources required.

We would be remiss if we did not discuss at least briefly the performance
characteristics of our code. As part of our calibration and testing,
we have performed detailed profiling studies, which we will not go
into detail here. Suffice to say that in its current configuration,
the code performs at approximately 20\% of peak on the Cray XT3. Its
weak scaling is linear (that is, its performance solving progressively
larger configurations, while keeping the load per processor constant,
scales linearly with the number of cores), while running on up to 4056
CPU cores on the Cray XT3 at PSC.

\section{Concluding remarks and outlook of future work}
\label{sec:conclusions}

We have presented a new computational framework (LEO), which we can use
to perform large--scale, high--resolution calculations in the context
of the characteristic approach in numerical relativity. This highly
parallel and easily extensible implementation has been used to solve the
model problem of a massless scalar field minimally coupled to gravity
(the three--dimensional Einstein--Klein--Gordon problem). We have shown
that the nonlinear code thus implemented is globally second-order convergent,
and how accurately we can follow quasi-normal mode ringing. We have studied 
the balance of energy for a number of initial data sets with different
angular structure. Aside from the interesting result of energy flow
saturation through the Schwarzschild horizon, the LEO framework offers a
good prospect to study new configurations beyond the linear regime and
the grid sizes used in this work.

Future directions we are currently exploring include the application
of the LEO framework to a consistent, quasilinear, fully first--order
formalism derived from~\cite{quasi}, and the extension of the model
problem considered here to massive scalar fields. The later case is
particularly important because it will allow us to simulate a
boson star orbiting a black hole. We will compare the performance of the
characteristic framework in ingoing versus outgoing null coordinates in
the extraction of quasi-normal modes, and in the study of nonlinear
effects in the neighborhood of the central black hole. The underlying
framework can be applied equally well to $3+1$ formulations of the Einstein
equations in spherical coordinates, in particular to a generalization
to three dimensions of the Bondi--Sachs gauge of~\cite{bsg}, and finally
to matched $3+1$ and characteristic evolutions~\cite{match1,match2,marsa}.

We have not addressed in the present work some outstanding problems with
the calculation of the News~\cite{bishop}, some of which arise from second
angular derivatives of the metric fields at $\mathscr{I}$ entering in
the computation of the News, a feature which can lead to substantial
propagation of errors. We have observed during our simulations that
the metric fields computed at $\mathscr{I}$ are smooth, as evident in
Figs.~\ref{fig:phi@scri}-\ref{fig:W@scri} (and so are those fields which
represent their angular derivatives, although these are not shown here).
It should be noted that, in our formulation~\cite{reduced}, the first
angular derivatives of some fields have been promoted to auxiliary
variables, for which a hypersurface equation is integrated radially.
(We do this for those fields whose second angular derivatives enter in
the computation of the News). In practice, this means that only first
order angular derivatives of any of the fields we evolve need to be
computed to calculate the news. It is possible that the cubed-sphere
approach leads to substantial improvements in the computed waveforms,
and this issue remains to be addressed. Although of potential importance
for the accuracy of gravitational waveforms, such a study lies outside
the scope of the present work.

\begin{acknowledgments}

R.G. wishes to thank Raghurama Reddy for many enlightening discussions
on the parallelization of the characteristic code, and acknowledges
the hospitality of the Universidad de Los Andes, M\'erida, Venezuela,
and the Kavli Institute for Theoretical Physics at Santa Barbara, where
portions of this project were carried out. W.B. thanks the Pittsburgh
Supercomputing Center for hospitality. This work was supported in part
by the National Science Foundation under grants No. PHY--0135390 to
Carnegie Mellon University, and No. PHY--0244752 and PHY-0555218 to
Duquesne University; by FONACIT under grants S1--98003270 and
F2002000426, and by CDCHT--ULA under grant C--1267--04--05--A. Code
development and simulations were carried out on the Cray XT3 at
the Pittsburgh Supercomputing Center, under grants PHY060004P and
PHY070022N. Additional computer time was provided by the Centro Nacional
de C\'alculo Cient\'{\i}fico, Universidad de Los Andes (CeCalcULA).

\end{acknowledgments}


\end{document}